\documentclass[onecolumn,secnumarabic,amssymb,superscriptaddress,nobibnotes,aps,pre,11pt]{revtex4-1}
\usepackage[toc,page,title,titletoc,header]{appendix}
\usepackage{stmaryrd}
\usepackage{amsmath}
\usepackage{tabularx}
\usepackage{subfigure}
\usepackage{graphicx}
\usepackage{epsfig}
\usepackage{color}

\newcommand{\abs}[1]{\lvert#1\rvert}

\begin{document}

\title{Sharp interface model for solid-state dewetting problems with weakly anisotropic surface energies}
\author{Yan Wang}
\affiliation{Department of Mathematics, National University of
Singapore, Singapore, 119076}

\author{Wei Jiang}
\email{jiangwei1007@whu.edu.cn}
\affiliation{School of Mathematics and statistics,
Wuhan University, Wuhan 430072, China}

\author{Weizhu Bao}
\affiliation{Department of Mathematics, National University of
Singapore, Singapore, 119076}
\affiliation{Center for Computational Science and Engineering, National University of Singapore, Singapore, 119076}

\author{David J. Srolovitz}
\affiliation{Departments of Materials Science and Engineering {\rm \&} Mechanical Engineering
and Applied Mechanics, University of Pennsylvania, Philadelphia, PA 19104, USA}


\begin{abstract}

We propose a sharp interface model for simulating solid-state dewetting where the surface energy is (weakly) anisotropic. The morphology evolution of thin films is governed by surface diffusion and contact line migration. The mathematical model is based on an energy variational approach. Anisotropic surface energies lead to multiple solutions of the contact angle equation at contact points. Introduction of a finite contact point mobility is both physically based and leads to robust, unambiguous determination of the contact angles. We implement the mathematical model in an explicit finite difference scheme with cubic spline interpolation for evolving marker points. Following validation of the mathematical and numerical approaches, we simulate the evolution of thin film islands, semi-infinite films, and films with  holes as a function of film dimensions, Young's angle $\theta_i$, anisotropy strength and crystal symmetry, and film crystal orientation relative to the substrate normal. We find that the contact point retraction rate can be well described by a power-law,  $l \sim t^n$. Our results demonstrate that the exponent $n$ is not universal -- it is sensitive to the Young's angle $\theta_i$ (and insensitive to anisotropy).  In addition to classical wetting (where holes in a film heal) and dewetting (where holes in a film grow), we observe cases where a hole through the film heals but  leave a finite size hole/bubble between the continuous film and substrate or where the hole heals leaving a continuous film that is not bonded to the substrate. Surface energy anisotropy  (i) increases the instability that leads to island break-up into multiple islands, (ii) enhances hole healing, and (iii) leads to finite island size even under some conditions where the Young's angle $\theta_i$ suggests that the film wets the substrate. The numerical results presented in the paper capture many of the complexities associated with solid-state dewetting experiments.

\end{abstract}
\date{\today}
\maketitle

\section{Introduction}

Solid-state dewetting of thin films on substrates has been  observed in a wide range of systems and is of considerable technological interest~\cite{Thompson12,Jiran90,Jiran92,Ye10a,Ye10b,Ye11a,Ye11b}. Unlike dewetting  of liquids  on substrates, this type of capillarity-driven dewetting occurs primarily through surface diffusion-controlled mass transport at temperatures well below the melting point of the film~\cite{Thompson12}. In a recent set of experiments, Ye and Thompson~\cite{Ye10a,Ye10b,Ye11a,Ye11b} demonstrated the geometric complexity and importance of crystalline anisotropy in dewetting. These, and related, recent experiments have led to renewed interest in understanding thin film dewetting and the influence of crystalline anisotropy on dewetting phenomena~\cite{Jiang12,Kim13,Zucker13,Rabkin11,Rabkin14a,Rabkin14b,Pierre11}.

The evolution of the morphology and rate of solid-state dewetting can be modeled as a type of surface-tracking problem. In addition to being a surface diffusion-mediated mass transport surface-tracking problem, it has the additional feature of a moving contact line. More specifically, the contact line is a triple line (where the film, substrate and vapor phases meet) that migrates as the surface evolves. Although  moving contact line problems have been extensively studied in the fluid mechanics community~\cite{Qian06,Ren07,Ren10,Ren11}, surface diffusion-mediated matter transport combined with moving contact lines pose a considerable challenge (it is a fourth-order partial differential equation  with moving boundaries ) for materials science, applied mathematics, and scientific computing.

Under the assumption that surface energies are isotropic, a mathematical model of  solid-state dewetting was first proposed by Srolovitz and Safran~\cite{Srolovitz86}. The model can be described in the following Lagrangian representation in two dimensions
~\cite{Srolovitz86,Wong00,Dornel06}:
\begin{equation}\label{eq1}
  \begin{cases}
    \displaystyle\frac{\partial\mathbf{X}}{\partial t} = V_n \mathbf{n},\\
    \displaystyle V_n = B\frac{\partial^2 \mu}{\partial s^2} = B \gamma_{\scriptscriptstyle {FV}}
    \frac{\partial^2 \kappa}{\partial s^2},
  \end{cases}
\end{equation}
where $\mathbf{X} = \big(x(s,t), y(s, t)\big)$ represents the moving film front (film/vapor interface) with arc length $s$ and time $t$,
$V_n$ is the moving velocity of the interface in the direction of its outward normal, $\mathbf{n}=(n_1,n_2)$ is the interface
outer unit normal direction, the chemical potential $\mu=\gamma_{\scriptscriptstyle {FV}}\kappa$, where $\gamma_{\scriptscriptstyle {FV}}$ is the surface energy density which is assumed as isotropic (a constant) in the model,
and $\kappa = \partial_{ss}x\, \partial_s y -\partial_{ss} y\,\partial_s x$ is the curvature of the interface. The material constant $B=D_s\nu\Omega^2/k_BT_e$, where $D_s$ is the surface diffusivity, $\nu$ is the number of diffusing
atoms per unit area, $\Omega$ is the atomic volume, $k_BT_e$ is the thermal energy.
Because the evolution includes contact point migration, Srolovitz and Safran proposed the following three boundary conditions for moving contact lines~\cite{Srolovitz86}:
\begin{subequations}
  \begin{equation}\label{bd1}
    y(x_c,t)=0,
  \end{equation}
  \begin{equation}\label{bd2}
    \frac{{\partial y}/{\partial s}}{{\partial x}/{\partial s}}(x_c,t)=\tan \theta_i,
  \end{equation}
  \begin{equation}\label{bd3}
    \frac{\partial \kappa}{\partial s}(x_c,t)=0,
  \end{equation}
\end{subequations}
where $x_c$ represents the moving contact point where the film, substrate and vapor meet and $\theta_i$ represents the equilibrium contact angle given by the classical Young equation, i.e. $\cos\theta_i=(\gamma_{\scriptscriptstyle
{VS}}-\gamma_{\scriptscriptstyle {FS}})/\gamma_{\scriptscriptstyle
{FV}}$, where $\gamma_{\scriptscriptstyle {FV}}$,
$\gamma_{\scriptscriptstyle {FS}}$ and $\gamma_{\scriptscriptstyle
{VS}}$ are, respectively, the surface energy densities of the film/vapor, film/substrate, and vapor/substrate interfaces. Condition (\ref{bd1}) ensures that the contact points always move along the substrate, condition (\ref{bd2}) comes from the force balance at the contact points, and condition (\ref{bd3}) ensures that the total mass of the thin film is conserved, implying that there is no mass flux at the contact points.

Based on the above model, Wong, {\it et al.}~\cite{Wong00,Du10} designed a ``marker particle'' numerical scheme to study the two-dimensional retraction of a discontinuous film (a film with a step) and the evolution of a perturbed cylindrical wire on a substrate; their numerical experiments indicated that the retracting film edges forms a thickened ridge followed by a valley; with increasing time, the ridge grows in height and the valley sinks, eventually  touching the substrate and leading to pinch-off events. Dornel, {\it et al.}~\cite{Dornel06} developed another numerical scheme to study the two main parameters in this problem, the film aspect ratio and the adhesion energy between the film and substrate, and quantified the retraction rate, breaking time and the number of islands formed. Jiang, {\it et al.}~\cite{Jiang12} developed a phase field method for simulating solid-state dewetting that naturally captures the topological changes that occur during evolution and that can be easily extended to three dimensions, avoiding the shortcomings of traditional front-tracking methods.

These earlier studies were based upon the assumption that all interface energies are isotropic. On the other hand, recent experiments have demonstrated that even in the case of cubic metals, crystalline anisotropy can strongly influence dewetting. More recently, Zucker, {\it et al.}~\cite{Kim13,Zucker13} analysed the important role played by  anisotropy during solid-state dewetting based on the crystalline method (Carter, {\it et al.}~\cite{Carter95}) applied to a two-dimensional model of edge retraction for highly anisotropic, full-faceted thin films. However, unlike in the isotropic case, valleys do not form ahead of the retracting ridge and hence pinch-off events do not occur in this model.  They also did not give an explicit governing equation and boundary conditions for the dewetting. In this paper, we develop a sharp-interface model for solid-state dewetting of thin films with weakly anisotropic surface energies.

This paper is organized as follows. First, we  present our new sharp interface model for simulating solid-state dewetting with  anisotropic interface energies including morphology evolution via anisotropic surface diffusion and contact line migration. We then propose a numerical scheme based on an explicit finite difference method combined with cubic spline interpolation for the evolving marker-particle points. Next, we perform a series of numerical tests, including numerical stability, relaxation and convergence issues. Finally, we apply these new results to simulate the morphology evolution of small and large islands on substrates, the retraction and pinch-off of semi-infinite films, and the evolution of films with holes with weakly anisotropic surface energies.

\section{Model formulation}
\begin{figure}[ht]
\centering
\includegraphics[width=10.0cm]{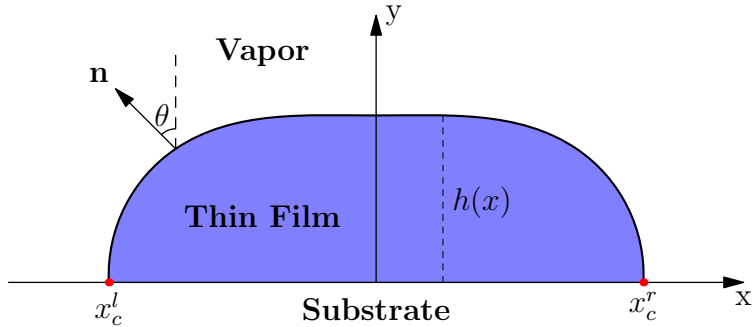}
\caption{A schematic illustration of a discontinuous solid thin film  on a flat, rigid substrate in two dimensions.
As the film morphology evolves, the contact points $x_c^l$ and $x_c^r$ move.}
\label{fig:2Dsetup}
\end{figure}

We first present a brief description of the sharp interface model with weakly anisotropic surface energy. Consider the case of a thin solid island  on a flat, rigid substrate in two dimensions, as illustrated in Fig.~\ref{fig:2Dsetup}. The total free energy of the system can be written as:
\begin{equation}
W=\int_{x_c^l}^{x_c^r}\left[\widetilde{\gamma}(\theta)(1+h_x^2)^{1/2}+
\gamma_{\scriptscriptstyle {FS}}-\gamma_{\scriptscriptstyle {VS}}\right]\;dx,
\label{eqn:2Denergy}
\end{equation}
where~$h=h(x)$~represents the thin film height \footnote{For simplicity of presentation, we assume $h(x)$ is a single-valued function of $x$. When $h(x)$ is a multi-valued function, this procedure is applied using an arc length parameterization.}, $x$ is the horizontal coordinate,
$h_x={\rm d} h/ {\rm d} x$, $\theta$ denotes the interface normal angle related to the film slope by $\cos \theta = 1/(1+h_x^2)^{1/2}$ (see Fig.~\ref{fig:2Dsetup}).  We assume that the film/vapor interface energy (density) is a function only of the normal angle, i.e. $\gamma_{\scriptscriptstyle {FV}}=\widetilde{\gamma}(\theta)$, and $x_c^l$ and $x_c^r$ represent the left and right moving contact points, respectively.

We calculate the first variation of the energy functional \eqref{eqn:2Denergy} with respect to the height function $h(x)$  and moving contact points ($x_c^l$ and $x_c^r$) as (see Appendix A for a more detailed derivation)
\begin{equation}
\frac{\delta W}{\delta h}=\Big(\widetilde{\gamma}(\theta)+\widetilde{\gamma}\,''(\theta)\Big)\kappa, \qquad x \in (x_c^l, x_c^r),
\label{eqn:variation4}
\end{equation}

\begin{equation}
\frac{\delta W}{\delta x_c^r}=\left[\widetilde{\gamma}(\theta)\cos\theta-\widetilde{\gamma}\,'(\theta)\sin\theta+\gamma_{\scriptscriptstyle {FS}}-
\gamma_{\scriptscriptstyle {VS}}\right]_{x=x_c^r},
\label{eqn:variation5}
\end{equation}

\begin{equation}
\frac{\delta W}{\delta x_c^l}=-\left[\widetilde{\gamma}(\theta)\cos\theta-\widetilde{\gamma}\,'(\theta)\sin\theta+\gamma_{\scriptscriptstyle {FS}}-
\gamma_{\scriptscriptstyle {VS}}\right]_{x=x_c^l}.
\label{eqn:variation6}
\end{equation}
Equation~\eqref{eqn:variation4} is the well known anisotropic Gibbs-Thomson relation. We  define the chemical potential of the system as $\mu=(\widetilde{\gamma}+\widetilde{\gamma}\,'')\kappa$. The term $\widetilde{\gamma}+\widetilde{\gamma}\,''=\Pi(\widetilde{\gamma})$ in the chemical potential, plays an important role in capilarity-driven morphology evolution. Spontaneous faceting can occur when $\Pi(\widetilde{\gamma})$ becomes negative, as pointed out in \cite{Eggleston01}. We  classify the anisotropy according to the value of $\Pi(\widetilde{\gamma})$.  The system is weakly anisotropic where $\Pi(\widetilde{\gamma})>0$ for all surface normal angles $\theta$; in this case, the surface is always smooth during the evolution and the anisotropic surface diffusion equation \eqref{eqn:variation4} is mathematically well-posed. The other class is strong anisotropy which occurs where $\Pi(\widetilde{\gamma})<0$ for some ranges of orientation angles $\theta$. In this case, some high energy surface orientations do not occur, such surfaces  undergo spontaneous faceting, Eq.~(\ref{eqn:variation4}) becomes ill-posed, and its solution is unstable. In this paper, we mainly focus on the weakly anisotropic case. Eqs.~\eqref{eqn:variation5} and~\eqref{eqn:variation6} can be used to form the contact angle boundary conditions by the following two approaches.

In the first approach, the force balance condition at the triple points is assumed to be applicable at all times, i.e. $\frac{\delta W}{\delta x_c^r}=0$ and $\frac{\delta W}{\delta x_c^l}=0$. The contact angle boundary conditions at triple points become~\cite{Min06}:
\begin{equation}
\widetilde{\gamma}(\theta)\cos\theta-\widetilde{\gamma}\,'(\theta)\sin\theta+\gamma_{\scriptscriptstyle {FS}}-
\gamma_{\scriptscriptstyle {VS}}=0.
\label{eqn:forcebalance}
\end{equation}
If the film/vapor interfacial energy is isotropic (i.e., $\widetilde{\gamma}$ is independent of $\theta$),
then Eq.~\eqref{eqn:forcebalance} reduces to the well-known isotropic Young equation. If the interfacial
energy is anisotropic (i.e. $\widetilde{\gamma}=\widetilde{\gamma}(\theta)$), a bending force $\widetilde{\gamma}\,'(\theta)$ appears which acts perpendicular to the film surface.
We refer to Eq.~\eqref{eqn:forcebalance} as the anisotropic Young equation.

\begin{figure}[ht]
\centering
\includegraphics[width=10.0cm]{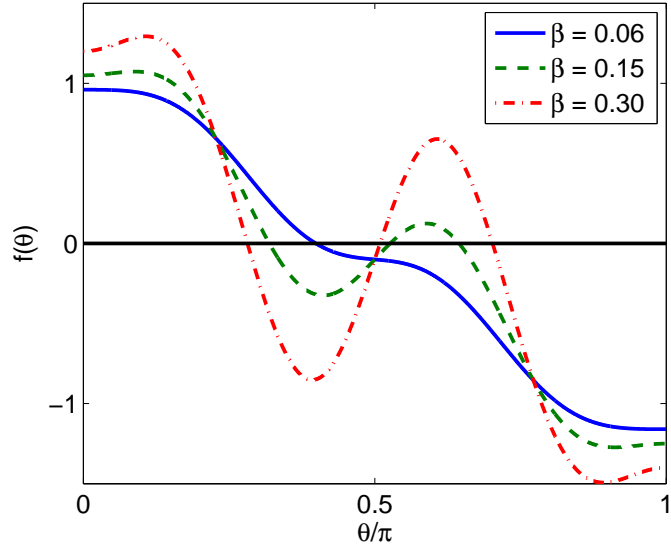}
\caption{The function $f(\theta)$  versus $\theta$ for  $0\le\theta\le\pi$, where $(\gamma_{\scriptscriptstyle {FS}}-\gamma_{\scriptscriptstyle {VS}})/\gamma_0=-0.1$.}
\label{fig:multisolution}
\end{figure}

This approach for determining the contact angle boundary conditions during the evolution requires the existence of a unique solution to Eq.~\eqref{eqn:forcebalance}. This is not always guaranteed for anisotropic surface energies. For example, consider the case where we write the  surface energy in the following form
\begin{equation}
 \widetilde{\gamma}(\theta) = \gamma_0(1+\beta\cos 4\theta),
\label{eqn:fourfold}
\end{equation}
for a crystal that posses four-fold rotation symmetry. Here $\gamma_0$ is a surface energy reference constant and the degree of anisotropy is controlled by the dimensionless coefficient $\beta$. When $\beta=0$, the system is isotropic. Increasing $\beta$ makes the system increasingly anisotropic. To observe the number change of the roots of the wetting angle equation \eqref{eqn:forcebalance}, as a function of  $\beta$, it is convenient to define
\begin{equation}
f(\theta)=\gamma(\theta)\cos\theta-\gamma\,'(\theta)\sin\theta+
\frac{\gamma_{\scriptscriptstyle {FS}}-\gamma_{\scriptscriptstyle {VS}}}{\gamma_0},
\label{eqn:rootnumber}
\end{equation}
where $\gamma(\theta)$ is the dimensionless anisotropic film/vapor interfacial energy,
$\gamma(\theta) = \widetilde{\gamma}(\theta)/\gamma_0$.  Figure~\ref{fig:multisolution} shows that as $\beta$ increases from $0.06$ to $0.30$, the number of roots of the wetting angle equation \eqref{eqn:forcebalance} changes from one to three.

To address the problem of multiple roots, we  use an alternative approach to determine the contact angle boundary conditions based upon a relaxation process at the triple points during the evolution.  More precisely, the velocity of the contact points is proportional to the magnitude of a relaxation force, i.e.,
\begin{equation}
 \frac{d x_c^r(t)}{d t}=-\eta\frac{\delta W}{\delta x_c^r}, \quad \text{at} \quad x=x_c^r,
\label{eqn:relaxationright}
\end{equation}

\begin{equation}
 \frac{d x_c^l(t)}{d t}=-\eta\frac{\delta W}{\delta x_c^l}, \quad \text{at} \quad x=x_c^l,
\label{eqn:relaxationleft}
\end{equation}
where $\delta W/\delta x_c^r$ and $\delta W/\delta x_c^l$ are given by Eqs. \eqref{eqn:variation5} and \eqref{eqn:variation6}, respectively, and $0<\eta\le \infty$ is a relaxation (or dissipative) coefficient.
When $\eta=\infty$, Eqs. (2.8)-(2.9) collapse to the  anisotropic Young equation (2.5).
With this assumption, the two-dimensional solid-state dewetting of a thin film on a solid substrate can be described in the following form by the sharp-interface model:
\begin{equation}
    \frac{\partial{\mathbf{X}}}{\partial t}=V_n \mathbf{n} = B\frac{\partial^2 \mu}{\partial s^2} \mathbf{n}=B\gamma_0\frac{\partial^2}{\partial s^2}
    \left[\Big(\gamma(\theta)+\gamma\,''(\theta)\Big)\kappa\right] \mathbf{n},
\label{governing_weak}
\end{equation}
where
$\gamma$ represents the dimensionless anisotropic film/vapor interfacial energy, i.e. $\gamma=\widetilde{\gamma}/\gamma_0$, and $B$ represents the material constant. For simplicity, we set the constant $B\gamma_0=1$ in the remainder of the paper.

The governing equation \eqref{governing_weak} for the solid-state dewetting problem is subject to the following conditions:
\begin{itemize}
\item[(i)] Contact point condition ({\bf {BC1}})
\begin{equation}
y(x_c^l,t)=0, \qquad y(x_c^r,t)=0,
\label{eqn:BC1}
\end{equation}
\item[(ii)] Relaxed (or dissipative) contact angle condition ({\bf {BC2}})
\begin{equation}\label{eqn:BC2a}
\frac{d x_c^l}{d t}=\eta\Big[\gamma(\theta_d^l)\cos\theta_d^l-\gamma\,'(\theta_d^l)
\sin\theta_d^l-\cos\theta_i\Big]_{x = x_c^l},
\end{equation}
\begin{equation}\label{eqn:BC2b}
\frac{d x_c^r}{d t}= - \eta\Big[\gamma(\theta_d^r)\cos\theta_d^r-\gamma\,'(\theta_d^r)
\sin\theta_d^r-\cos\theta_i\Big]_{x = x_c^r},
\end{equation}
where $\theta_i$ is defined as the corresponding isotropic Young contact angle, i.e. $\cos \theta_i = (\gamma_{\scriptscriptstyle {VS}}-\gamma_{\scriptscriptstyle {FS}})/\gamma_0$, $\theta_d^l$ and $\theta_d^r$ are the (dynamic) contact angle at the left and right contact points, respectively.
\item[(iii)] Zero-mass flux condition ({\bf {BC3}})
\begin{equation}
\frac{\partial \mu}{\partial s}(x_c^l,t)=0, \qquad \frac{\partial \mu}{\partial s}(x_c^r,t)=0,
\label{eqn:BC3}
\end{equation}
and this condition is necessary for the total mass (denoted as $A(t)$) conservation of the thin film.
\end{itemize}

The introduction of relaxation kinetics for the contact point position, Eqs. \eqref{eqn:BC2a} and \eqref{eqn:BC2b}, is not simply a  tool for stabilizing the numerical method. It has its origin in the complex atomic structure of the contact point, where typically atoms are not all exactly on perfect crystal sites. This variation in the atomic structure in the vicinity of the contact point can be associated with elastic deformation, slipping between film and substrate, dislocations at the film/substrate interface, reconstruction of the interfaces, and other forms of non-elastic deformation.  The local distortion of the atomic lattice at the contact point must be propagated along with the moving contact point and because its structure is distinct from that of the remaining film or film/substrate interface it has its own distinct kinetics.  Hence, we can think of this contact point as having a unique mobility $M_c=\eta$.  A similar concept was introduced to describe the effect of grain boundary triple junctions (where three grain boundaries meet) on the motion of grain boundaries (see e.g., \cite{Czubayko98,Upmanyu02}) and contact lines in liquid film wetting of substrates (see e.g., \cite{Qian06,Ren07,deGennes85,Ralston08}).

\section{Numerical method}

The  governing equations \eqref{governing_weak}-\eqref{eqn:BC3} are solved by an explicit finite difference method combined with a cubic spline interpolation for evolving marker points. The detailed algorithm at the $k^\text{th}$ time step is as follows.

First, suppose that there are $N+1$ marker points uniformly distributed on the film/vapor interface (curve)
with respect to the arc length at the time step $k$. We denote the total arc length of the curve as $L^{k}$, the mesh size as $h^{k}:= L^{k}/N$, the time step as $\tau^{k}$, and the uniformly distributed marker points as $(x_j^k, y_j^k), j =0,1,\ldots, N$. Evolving the $N+1$ marker points according to Eqs.~\eqref{governing_weak}-\eqref{eqn:BC3}
based on the following explicit finite difference method,  we obtain the positions of the $N+1$ marker points at the time step $k+1$, denoted as $(\tilde{x}_j^{k+1}, \tilde{y}_j^{k+1}), j = 0,1,\ldots,N$. In addition, we denote $\kappa_j^k, \gamma_j^k$ and $\mu_j^k$ to be approximations to the curvature, the dimensionless film/vapor surface energy density, and the chemical potential on the $j^\text{th}$ marker point at the $k^\text{th}$ time step. Next, we introduce the following finite difference discretization operators:
\[
\delta_t^+ x_j^k = \frac{\tilde{x}_j^{k+1} - x_j^k}{\tau^{k}}, \quad \delta_s x_j^k = \frac{x_{j+1}^k - x_{j-1}^k}{2h^{k}},
\]
\[
\delta_s^2 x_j^k = \frac{x_{j+1}^k - 2x_j^k + x_{j-1}^k}{(h^{k})^2}.
\]
Using a central finite difference scheme for discretizing the spatial derivatives and a forward Euler scheme for discretizing the temporal derivatives, the governing equations \eqref{governing_weak} become
\begin{equation}\label{discrete}
  \begin{cases}
    \displaystyle\delta_t^+ x_j^k = - \delta_s^2 \mu_j^k\cdot \delta_s y_j^k,\\
    \displaystyle\delta_t^+ y_j^k = \delta_s^2 \mu_j^k\cdot \delta_s x_j^k,\\
    \displaystyle \mu_j^k = (\gamma_j^k + (\gamma\,'')_j^k)\kappa_j^k, \\
    \displaystyle\kappa_j^k = \delta_s y_j^k \cdot \delta_s^2 x_j^k - \delta_s x_j^k \cdot \delta_s^2 y_j^k,
  \end{cases}
  \quad j = 1,2,\ldots,N-1,
\end{equation}
and the boundary conditions \eqref{eqn:BC1}-\eqref{eqn:BC3} become
\begin{equation}
  \tilde{y}_0^{k+1} = \tilde{y}_{N}^{k+1} = 0,
\end{equation}
\begin{equation}
  \delta_t^+ x_0^k = \eta\left[\gamma_0^k \cos(\theta_0^k) - (\gamma\,')_0^k   \sin(\theta_0^k) - \cos\theta_i\right],
\end{equation}
\begin{equation}
  \delta_t^+ x_{N}^k = -\eta\left[\gamma_{N}^k \cos(\theta_{N}^k) - (\gamma\,')_{N}^k \sin(\theta_{N}^k) - \cos\theta_i\right],
\end{equation}
\begin{equation}
  \mu_0^k = \frac{4}{3}\mu_1^k - \frac{1}{3}\mu_2^k , \quad \mu_{N}^k = \frac{4}{3}\mu_{N-1}^k - \frac{1}{3}\mu_{N-2}^k.
\end{equation}

Based on this numerical scheme, we  immediately obtain the positions of the marker points $(\tilde{x}_j^{k+1}, \tilde{y}_j^{k+1}), j=0,1,\ldots,N$. Note that these marker points
may not be uniformly distributed along the curve
with respect to the arc length. Thus, we redistribute these marker points via a cubic spine interpolation such
that they are uniformly distributed as follows. First, making use of these new marker points $(\tilde{x}_j^{k+1}, \tilde{y}_j^{k+1}), j=0,1,\ldots,N$, we construct a piecewise curve $\{(X_j^{k+1}(p),  Y_j^{k+1}(p)), \; p\in [(j-1)h^{k}, jh^{k}]\}_{j=1,2,\ldots,N}$ by using a cubic spline interpolation. Here, $X_j^{k+1}(p)$ and $Y_j^{k+1}(p)$ are cubic polynomials obtained from a cubic spline interpolation for the points $\{(jh^{k}, \tilde{x}_j^{k+1}), j=0,1,\ldots,N\}$ and $\{(jh^{k}, \tilde{y}_j^{k+1}), j= 0,1,\ldots,N\}$, respectively. By using these cubic polynomials, we directly compute the arc length of each piecewise cubic polynomial curve, denoted as $L^{k+1}_j, j=1,2,\ldots,N$. Then, we obtain the total arc length $L^{k+1}=\sum\limits_{j=1}^N L_j^{k+1}$ and determine the uniform mesh size at the $(k+1)^\text{th}$ time step as $h^{k+1}=L^{k+1}/N$. In order to redistribute the $N+1$ points uniformly according
to the arc length for the $(k+1)^\text{th}$ time step computation, we set $x_0^{k+1}=\tilde{x}_0^{k+1}$, $y_0^{k+1}=\tilde{y}_0^{k+1}=0$, $x_N^{k+1}=\tilde{x}_N^{k+1}$, $y_N^{k+1}=\tilde{y}_N^{k+1}=0$.
For each fixed $j=1, 2,\ldots,N-1$, we first locate to which unique
piecewise cubic polynomial curve the new $j^\text{th}$ point $(x_j^{k+1},y_j^{k+1})$ belongs, i.e. finding a unique $1\le i\le N$ such that $\sum\limits_{l=1}^{i-1}L_l^{k+1}
\le j h^{k+1}<\sum\limits_{l=1}^{i} L_l^{k+1}$, then numerically solve the following equation
\begin{eqnarray*}
g(q)&=&\int_{(i-1)h^{k}}^{(i-1)h^{k}+q} \sqrt{\left(\frac{dX_{i}^{k+1}(p)}{dp}\right)^2 +
\left(\frac{dY_{i}^{k+1}(p)}{dp}\right)^2} \;{\rm d} p   \\
&=&jh^{k+1} - \sum_{l=1}^{i-1} L^{k+1}_{l},\quad
0\le q< h^k,
\end{eqnarray*}
to obtain its unique root $q=q^*$, and finally the position of the $j^\text{th}$ uniformly distributed marker point
at the $(k+1)^\text{th}$ time step is obtained as $x_j^{k+1}=X_i^{k+1}((i-1)h^{k}+q^*)$ and $y_j^{k+1}=Y_i^{k+1}((i-1)h^{k}+q^*)$.

\section{Numerical Results}

We now present the results from several simulations using the sharp interface method presented above to determine the effect of the relaxation coefficient/contact point viscosity $\eta$.
We then apply the  model and numerical algorithm to simulate  solid-state dewetting in several different thin film geometries with weakly anisotropic surface energy in  two dimensions. For simplicity, we set the initial film thickness to unity and assume an anisotropic surface energy of the form:
\begin{equation}
  \gamma(\theta) = 1+\beta\cos[m(\theta+\phi)],
  \label{eqn:sfenergy}
\end{equation}
where $\beta$ is the degree of anisotropy, $m$ is the order of the rotational symmetry and $\phi$ represents a phase shift angle describing a rotation of the crystallographic axes from a reference orientation (the substrate plane). The surface energy can be viewed as weakly anisotropic for $0\le \beta<\frac{1}{m^2-1}$. In this paper, $\phi$ is set to 0, except where noted. In addition, because the numerical scheme presented above is explicit, the time steps in our simulations are always chosen to be of $O((h^k)^4)$ to ensure numerical stability.

\subsection{Relaxation coefficient/contact point viscosity}

The relaxation coefficient/contact point viscosity $\eta$ determines the rate of relaxation of the dynamic contact angle $\theta_d$ to the equilibrium contact angle $\theta_a$ which satisfies the anisotropic Young equation (2.5). In general, for small $\eta$, the relaxation is very slow and the contact points move very slowly. On the other hand, if $\eta$ is very large, the relaxation process occurs very quickly such that the dynamic contact angle $\theta_d$  quickly converges to one of the solutions of the anisotropic Young equation $\theta_a$.  In this case, the time steps for
numerically integrating Eqs. (2.12)-(2.13)
must be chosen very small in order to maintain numerical stability. From the point of view of numerics, the choice of the relaxation coefficient must represent a balance between these factors. On the other hand, in any physical system, $\eta$ is a material parameter and must be determined either from experiment or microscopic (e.g., molecular dynamics) simulations.

\begin{figure}
\centering
\includegraphics[width=0.95\textwidth]{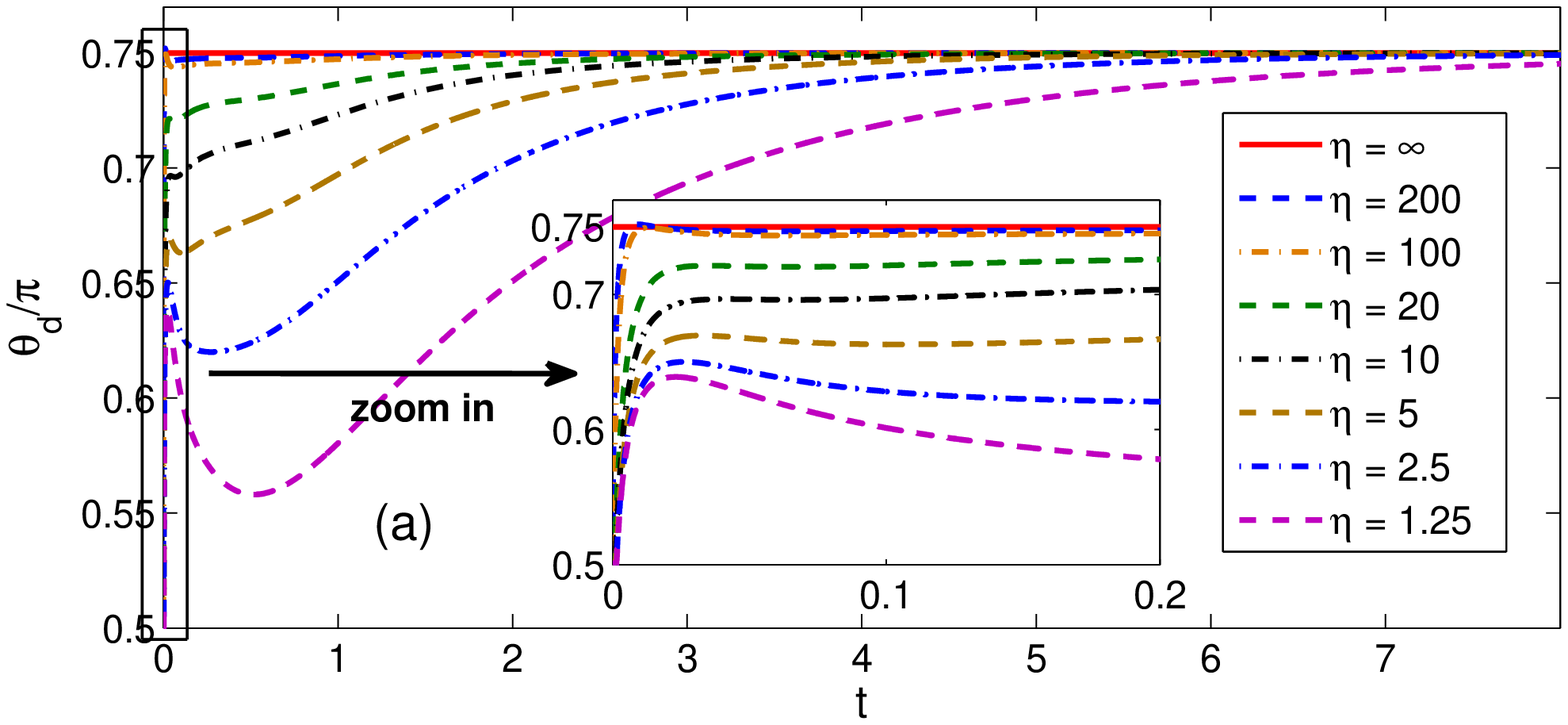}
\includegraphics[width=0.95\textwidth]{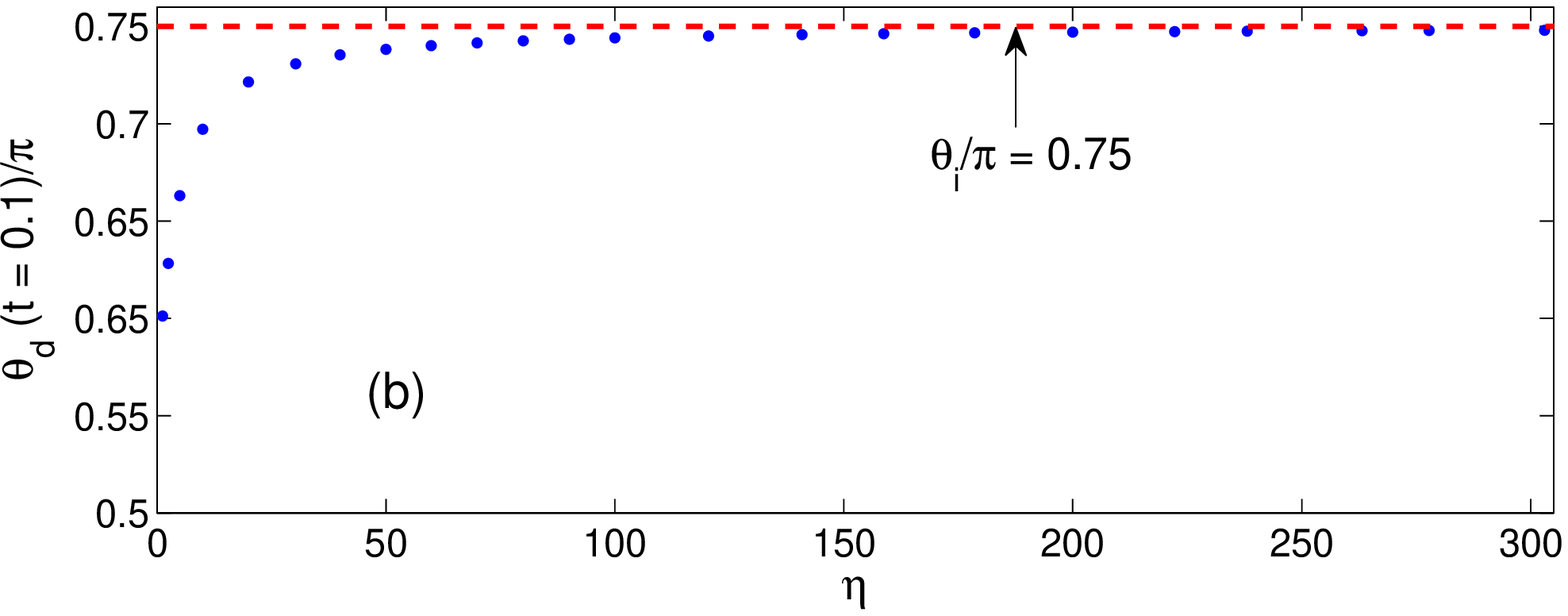}
\caption{(a) The dynamic contact angle $\theta_d$ as a function of time for several different relaxation coefficients and simulation parameters $\beta=0$, $\theta_a=\theta_i=3\pi/4\approx 2.36$ (upper panel).  The initial island is rectangular with   length $L=5$, thickness $h=1$ and $\theta_d(t=0)=\pi/2$. (b) The dynamic angle $\theta_d$ measured at $t = 0.1$ as a function of the contact point viscosity $\eta$.}
\label{fig:eta_ang}
\end{figure}

Figure~\ref{fig:eta_ang}a shows the evolution of the dynamic contact angle $\theta_d$ as a function of time for eight different relaxation coefficients ($\eta=1.25,2.5,5,10,20,100,200,\infty$) for a case of an initially long, thin rectangular island (length $L=5$, thickness $h=1$) with $\beta=0$ and $\theta_i=3\pi/4$.  The  contact angle, initially grows very quickly from its initial value of $\pi/2$ to a near steady-state dynamical value (see the inset to Fig.~\ref{fig:eta_ang}a).  Then, as the island approaches its equilibrium shape the contact point slows and $\theta_d \rightarrow \theta_i$ (note that $\theta_a=\theta_i$ in the isotropic case).  The near steady-state dynamical angle, seen in the inset for large $\eta$, is always smaller than the equilibrium value $\theta_i$ (the $\eta=\infty$ case) and $\theta_d$ increases with increasing contact point viscosity $\eta$ (see Fig.~\ref{fig:eta_ang}b).  This is consistent with experimental and atomistic simulation observations of the effect triple junction drag on dynamic triple junction angles in grain boundaries \cite{Czubayko98,Upmanyu02} and in contact lines of fluids on substrates~\cite{Ren07,Ren10}.

In order to further clarify the effects of the choice of the relaxation coefficient $\eta$, we performed a series of numerical simulations of the evolution of an initially rectangular, thin film island of three different initial lengths $L=5, 100$ and semi-infinite, for several values of $\eta$ and for $\beta=0$ and $\theta_i=3\pi/4$ and different coefficients. When $L=5$, the island evolves to an arc of a circle (equilibrium state) and the simulations are terminated when the maximum  error in the adjacent time level of marker point separation is smaller than a threshold value. For the $L=100$ and semi-infinite cases, the simulations are terminated when the first pinch-off event (the film thins to zero thickness creating new contact points) occurs. We compared the results for three different values of $\eta = 10,20,100$ and found that $\eta$ has no discernible effect on the equilibrium island shapes (not shown).  $\eta$  also had very little effect on the simulation termination/island equilibration times  (see Table~\ref{tab:TT}). For the semi-infinite thin film case, we numerically computed the contact point position as a function of time and found that it is  well described by a power law with the value $0.42$, regardless of the relaxation coefficient $\eta$~\cite{Thompson12,Kim13}. Unless otherwise noted, the simulations reported below were all performed with $\eta=100$.
\begin{table}
  \centering
  \begin{tabular}{c||c|c|c|c}
    \hline
    $\eta$ & 200 & 100 & 20 & 10 \\
    \hline
    $L = 5$ & $1.3445\times 10^{1}$ & $1.3488\times 10^{1}$ & $1.3842\times 10^{1}$ & $1.4309\times 10^{1}$ \\
    \hline
    $L = 100$  & $1.4055 \times 10^{3}$ & $1.4053 \times 10^{3}$ & $1.4095 \times 10^{3}$ & $1.4094 \times 10^{3}$ \\
    \hline
    semi-infinite  & $1.6397 \times 10^{4}$ & $1.6394 \times 10^{4}$ & $1.6392 \times 10^{4}$ & $1.6392 \times 10^{4}$ \\
    \hline
  \end{tabular}
  \caption{Equilibration times for rectangular islands of thickness $h=1$ and several initial lengths for different relaxation coefficients $\eta$ (see the text for more details).}
  \label{tab:TT}
\end{table}

\subsection{Convergence test}

\begin{figure}
\centering
\includegraphics[width=0.9\textwidth]{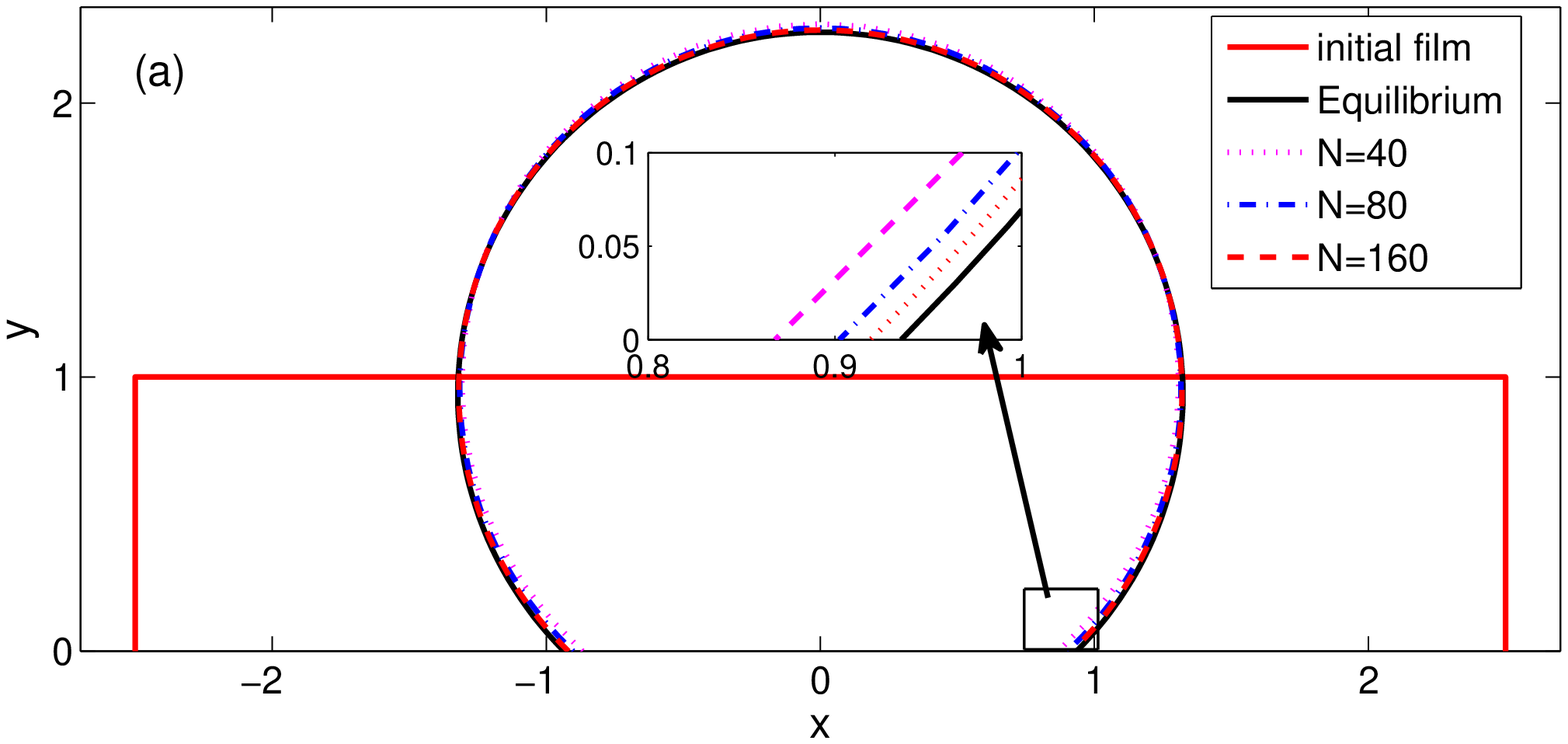}
\includegraphics[width=0.9\textwidth]{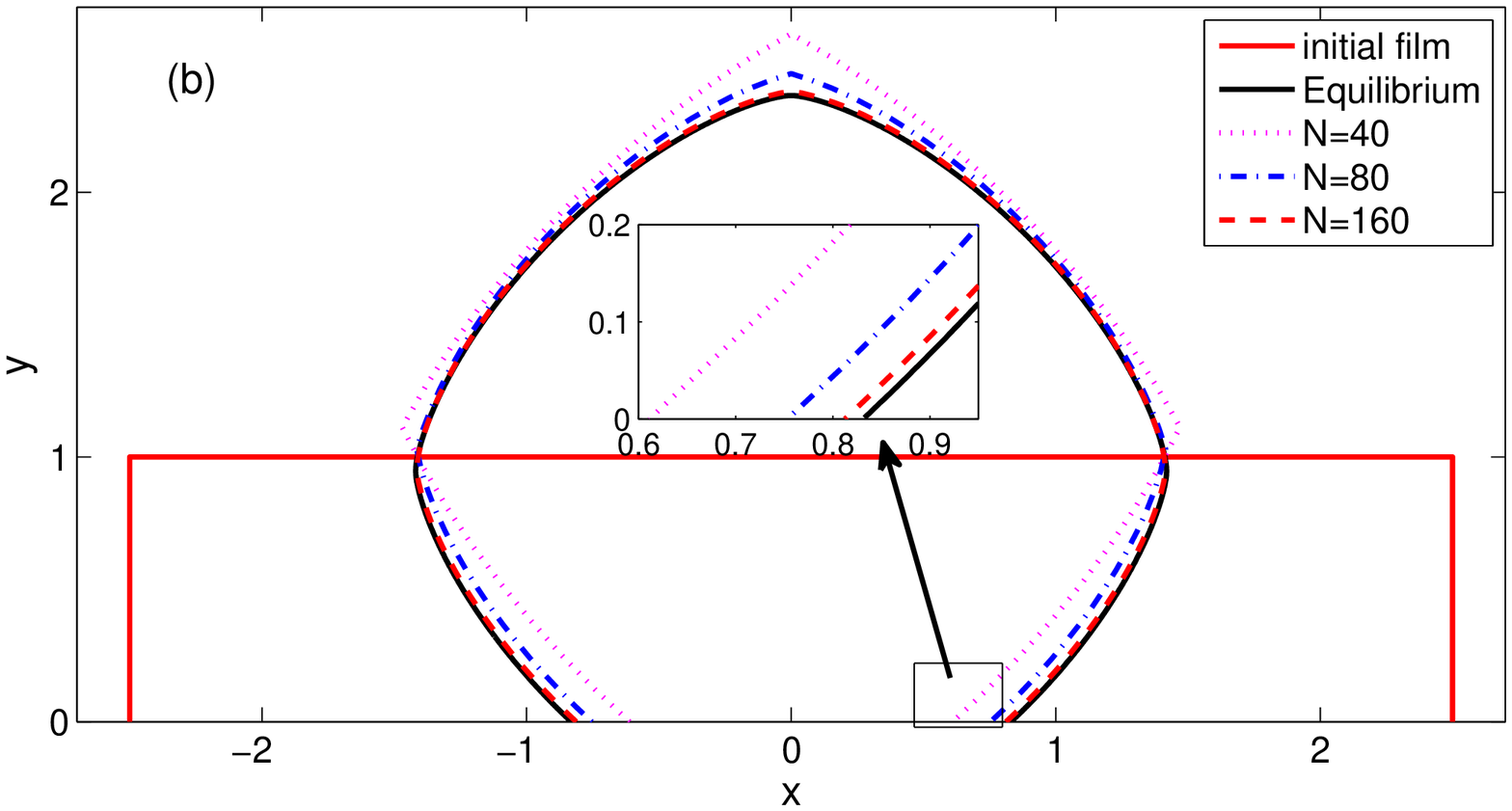}
\caption{Comparison of the long time numerical solution of the dynamic island shape with the theoretical equilibrium shape (from the Winterbottom construction, shown in blue) for several values of the number of computational marker points $N$ for: (a) the isotropic surface energy case with $\beta = 0$ and $\theta_i = 3\pi/4$; and (b) the weakly surface energy case with $\beta = 0.06$, $\theta_i = 3\pi/4$ and $m = 4$. }
\label{fig:convergence}
\end{figure}

We now investigate the convergence  of the numerical scheme by performing simulations for a rectangular island of length $L=5$ and thickness $h=1$. In this case, the  equilibrium island shape can be determined using the Winterbottom construction~\cite{winterbottom67} (see Appendix B for more details). We compare the numerical equilibrium island shape with the theoretical  predictions as a function of the number of markers $N$ employed in the description of the island shape. Figure~\ref{fig:convergence} and Table~\ref{tab:contactpt}  show the numerical convergence results.

\begin{table}[t]
  \centering
  \begin{tabular}{c||c|c|c||c|c|c}
    \hline
     & \multicolumn{3}{c||}{isotropic} & \multicolumn{3}{c}{anisotropic}\\
    \hline
    N & 40 & 80 & 160 & 40 & 80 & 160\\
    \hline
    $\alpha_{\rm err}$ & 0.0721 & 0.0354 & 0.0175 & 0.2644 & 0.0937 & 0.0227\\
    \hline
    $d_{\rm err}$ & 0.0675 & 0.0331 & 0.0163 & 0.2337 & 0.0832 & 0.0189\\
    \hline
  \end{tabular}
  \caption{Convergence of the long time simulation island shape with the theoretical equilibrium shape (Winterbottom construction) as a function of the number of marker points $N$. The error measures $\alpha_\text{err}$ and $d_\text{err}$ are defined in the text.}
  \label{tab:contactpt}
\end{table}

\begin{figure}
\centering
\includegraphics[width=0.8\textwidth]{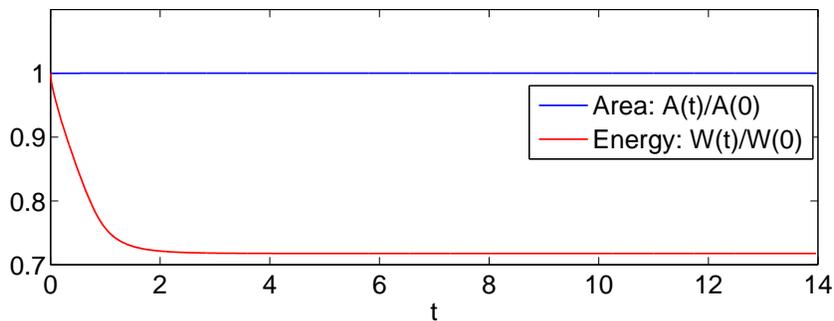}
\caption{The temporal evolution of the normalized total free energy and the normalized area occupied by the  island for the weakly anisotropic case with $N=80$ and $\beta = 0.06$ presented in Fig.~\ref{fig:convergence}b.}
\label{fig:energy}
\end{figure}

As showed in Fig.~\ref{fig:convergence}, the numerical equilibrium states converge to the theoretical equilibrium states (Winterbottom construction, shown by the black curves) with increasing number of marker points from $N=40$ to $N=160$ in both the isotropic and weakly anisotropic cases;  this is a clear demonstration of the convergence of our numerical scheme. We also computed the relative error $\alpha_{\rm err}$ of the right contact point position between the numerical equilibrium state $x_{c,n}^r$ and the theoretical equilibrium state $x_{c,e}^{r}$, and the maximum distance error
$d_{\rm err}$ between the two equilibrium shapes measured by marker points.  We define the relative error as $\alpha_{\rm err}=|(x_{c,n}^{r}-x_{c,e}^{r})/x_{c,e}^{r}|$. Table~\ref{tab:contactpt} shows the convergence of the numerical equilibrium shape to the theoretical equilibrium shape. From Table~\ref{tab:contactpt}, we see that the shapes are determined more accurately in the isotropic than in the anisotropic case for the same number of marker points. This can be understood by noting that in the anisotropic surface energy case, more marker points are required to capture the faceting morphology than in the smoother isotropic case. In addition, we also computed the temporal evolution of the normalized total free energy $\overline{W}(t)$ and the normalized island size (area) $\overline{A}(t)$ in the weakly anisotropic case shown in
Figure~\ref{fig:energy} which demonstrates that the area
occupied by the island is conserved (mass conservation) during the entire
simulation  and that the total free energy of the system decays monotonically during the evolution.

\section{Island/Film Evolution Simulation Results}

We now examine dewetting in several geometries using the mathematical model and numerical scheme described above for weakly anisotropic surface energies. First, we examine the evolution of small islands on a flat substrate with different degrees of anisotropy and $m$-fold crystal symmetries. Next, we perform numerical simulations for the evolution of large islands and semi-infinite films on a substrate, where pinch-off occurs. Then, we examine the relationship among the number of agglomerates resulting from the evolution of islands, the initial island size $L$ and the isotropic Young angle $\theta_i$. Finally, we examine the evolution of an infinite long thin film containing holes.

\subsection{Small islands}

\begin{figure}
\centering
\includegraphics[width=1\textwidth]{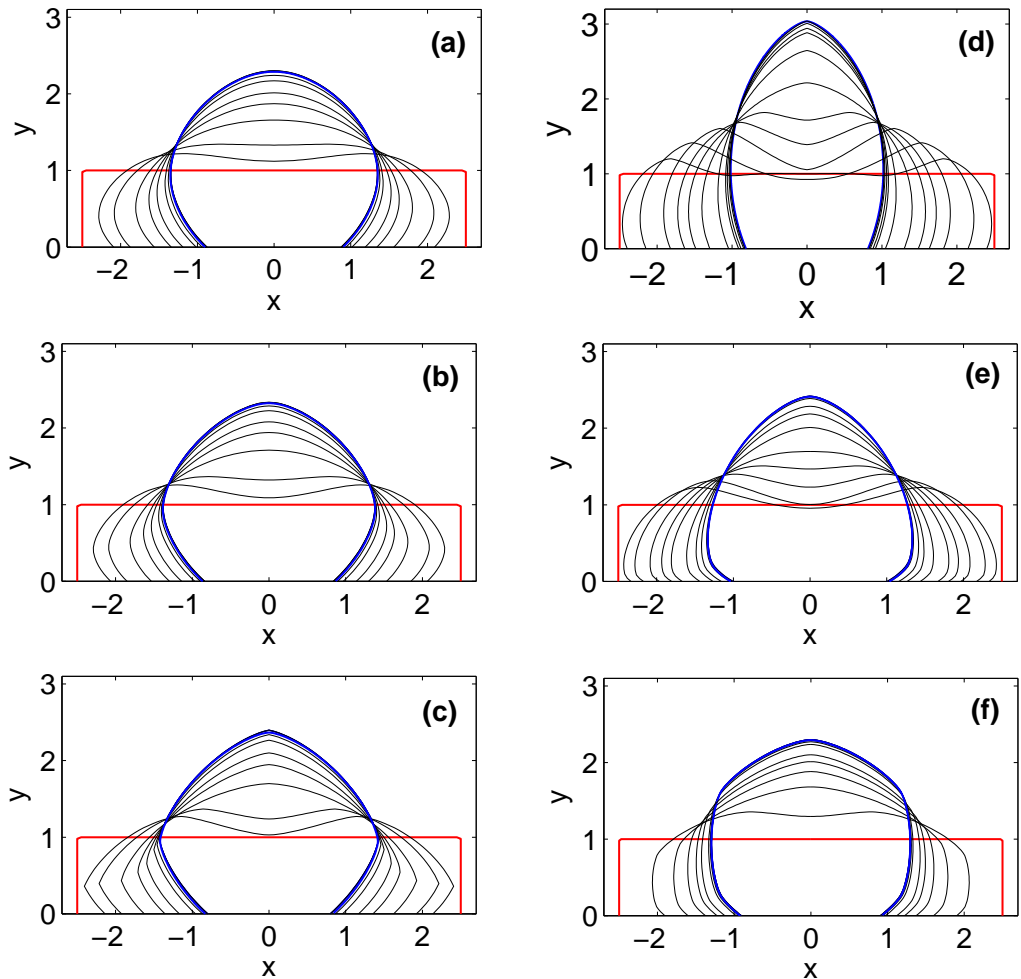}
\caption{Several steps in the evolution of small, initially rectangular islands (shown in red) toward their equilibrium (Winterbottom) shape (shown in blue) for different anisotropies $\beta$ and crystalline rotational symmetry orders $m$ ($\theta_i = 3\pi/4$ in all cases). Figures (a) - (c) are  results for $\beta = 0.02$, $0.04$, $0.06$ ($m = 4$ are fixed). Figures (d) - (f) are simulation results for (d) $m = 2$, $\beta = 0.32$, (e) $m = 3$, $\beta = 0.1$, and (f) $m = 6$, $\beta = 0.022$, respectively.}
\label{fig:sbeta}
\end{figure}

The evolution of small rectangular islands towards their equilibrium shapes is shown in Fig.~\ref{fig:sbeta} for several different anisotropy strengths $\beta$ and $m$-fold crystalline symmetries for fixed $\theta_i=3\pi/4$. In all cases, the dynamic contact angle $\theta_d$ rapidly converges to the equilibrium contact angle $\theta_a$ and then remains fixed throughout the remainder of the island shape evolution. As the anisotropy $\beta$ increases from $0.02$ to $0.06$ (Fig.~\ref{fig:sbeta}a--c), the equilibrium island shape changes from smooth and nearly circular to an increasingly faceting shape with increasingly sharp corners, as expected based upon the anisotropic surface energy. As the rotational symmetry $m$ (Fig.~\ref{fig:sbeta}d--f) is increased, the number of facets in the equilibrium shape increases.

\begin{figure}[ht]
  \centering
    \includegraphics[width=0.60\textwidth]{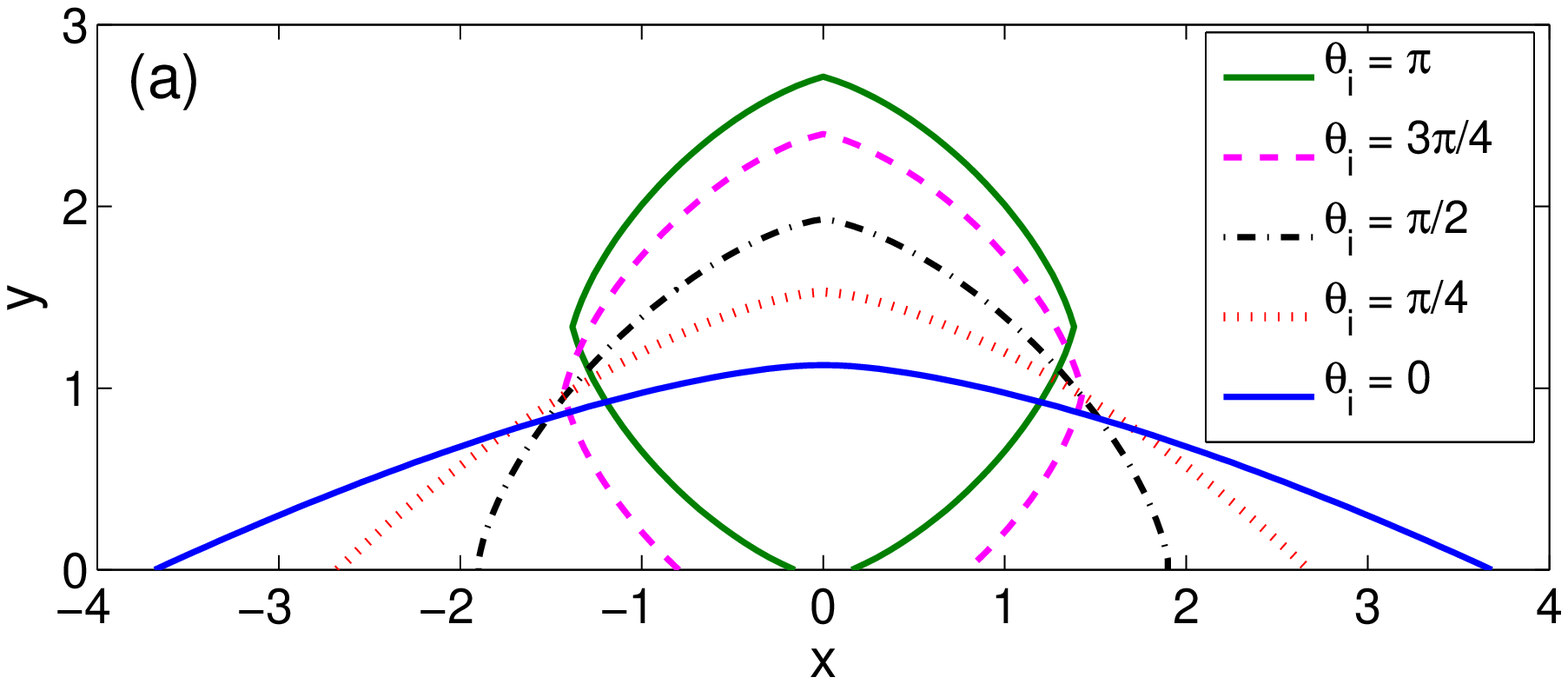}
    \includegraphics[width=0.37\textwidth]{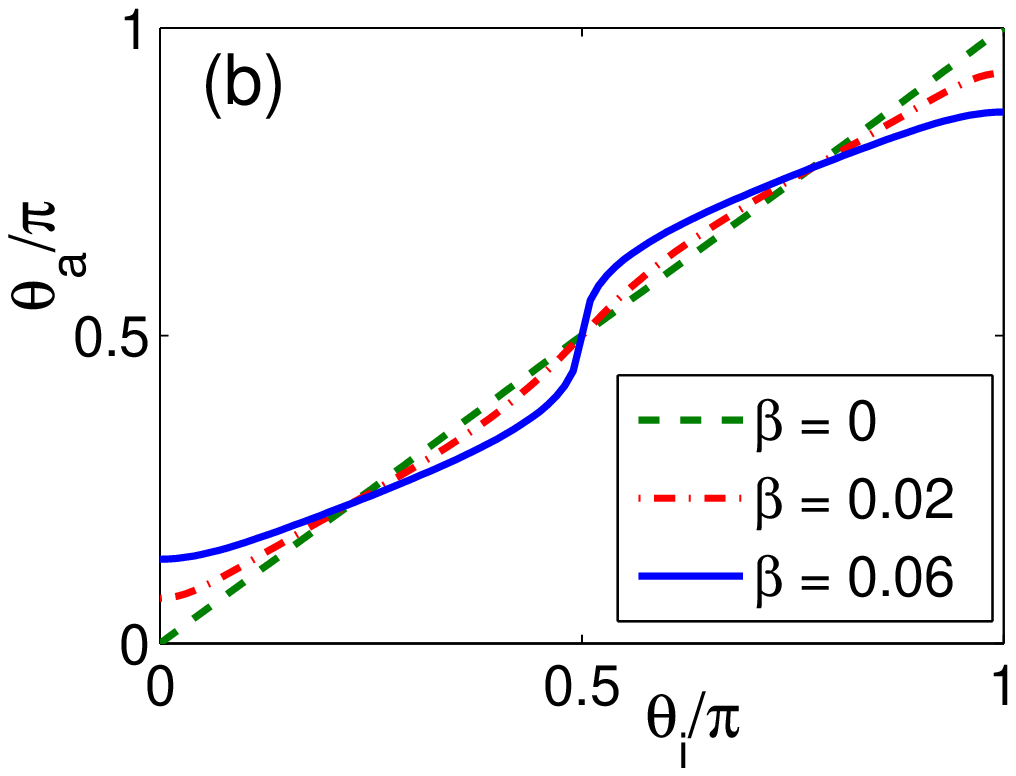}
    \caption{Equilibrium morphologies resulting from the evolution of several small $L=5$ islands. Figure (a) shows the results for different values of $\theta_i$ ($\beta = 0.06, m = 4$). Figure (b) shows the relationship between the anisotropic equilibrium contact angle $\theta_a$ and $\theta_i$ for different magnitude of anisotropies $\beta$.   }
\label{fig:sang}
\end{figure}

Fig.~\ref{fig:sang}a shows the equilibrium shapes of small islands (initially rectangular with $L=5$, $h=1$) for  different values of the Young angle $0\le\theta_i\le\pi$ for $\beta=0.06, m=4$. Unlike in the isotropic case (even though $\theta_i=0$ or $\pi$), complete wetting (or dewetting) does not occur with anisotropic surface energies. This can be understood by noting that the bending term which appears in the anisotropic Young equation \eqref{eqn:forcebalance} and is absent in its isotropic analogue, leads to an equilibrium angle $\theta_a$ that differs from $\theta_i$, and is not $0$ or $\pi$  even when $\theta_i=0$ or $\pi$ (Fig.~\ref{fig:sang}b).

\begin{figure}
\centering
  \includegraphics[width=0.655\textwidth]{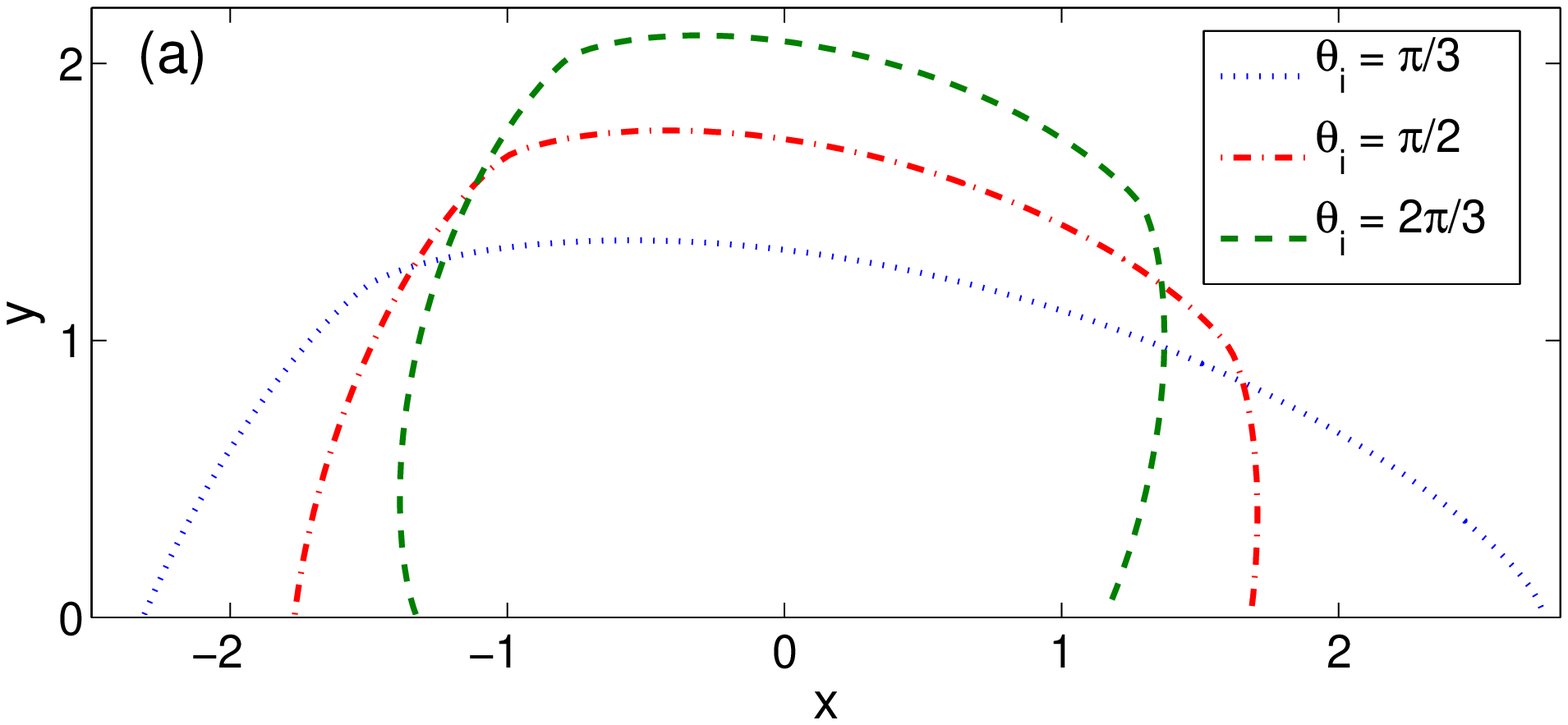}
  \includegraphics[width=0.335\textwidth]{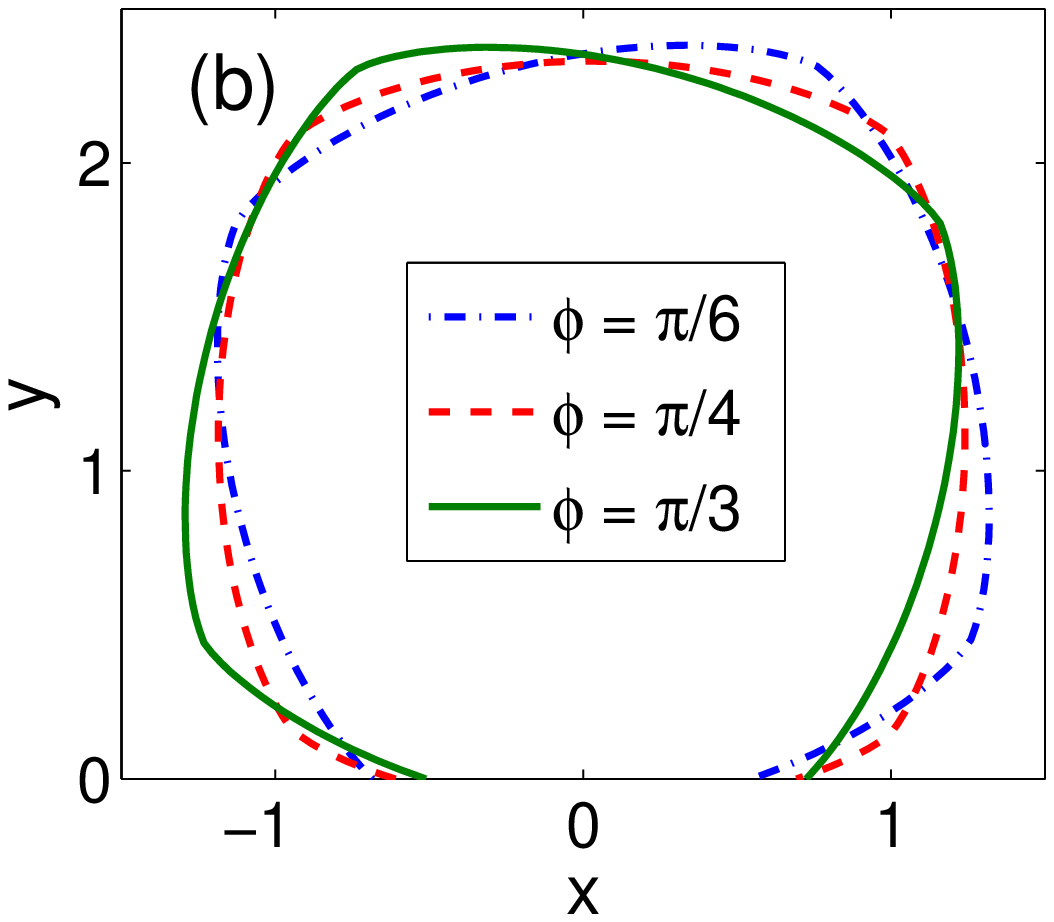}
\caption{(a) Equilibrium island morphologies for small  ($L = 5$) islands with a rotation of the crystal relative to the surface normal of $\phi = \pi/3$ for different values of $\theta_i$. (b) Equilibrium island morphologies for small  ($L = 5$) islands with $\theta_i = 5\pi/6$ for several different crystal rotations  $\phi$ (phase shifts).  In both figures,  $\beta=0.06$ and $m=4$.   }
\label{fig:non}
\end{figure}

We also performed numerical simulations of the evolution of small islands with finite values of $\phi$ in Eq.~\eqref{eqn:sfenergy}
for the weakly anisotropic cases for $\beta = 0.06, m = 4$ --- this corresponds to different rotations of the crystalline axis of the island relative to the substrate normal. The numerical equilibrium shapes for different $\theta_i$ and phase shift angles $\phi$ are shown in Figs.~\ref{fig:non}a and \ref{fig:non}b, respectively. The asymmetry of the equilibrium shapes is clearly seen in the two figures, resulting from breaking the symmetry of the surface energy anisotropy (see Eq.~\eqref{eqn:sfenergy}) with respect to the substrate normal. The numerical results confirm that the left and right equilibrium contact angles are two roots of the anisotropic Young equation \eqref{eqn:forcebalance}.  In general, it is possible for a crystal island with an $m$-fold rotation symmetry to exhibit $0$ to $m-1$ corners upon rotation of the crystal axes with respect to the substrate $\phi$ and the isotropic wetting angle $\theta_i$.

\subsection{Large islands}

\begin{figure}
\centering
  \includegraphics[width=0.9\textwidth]{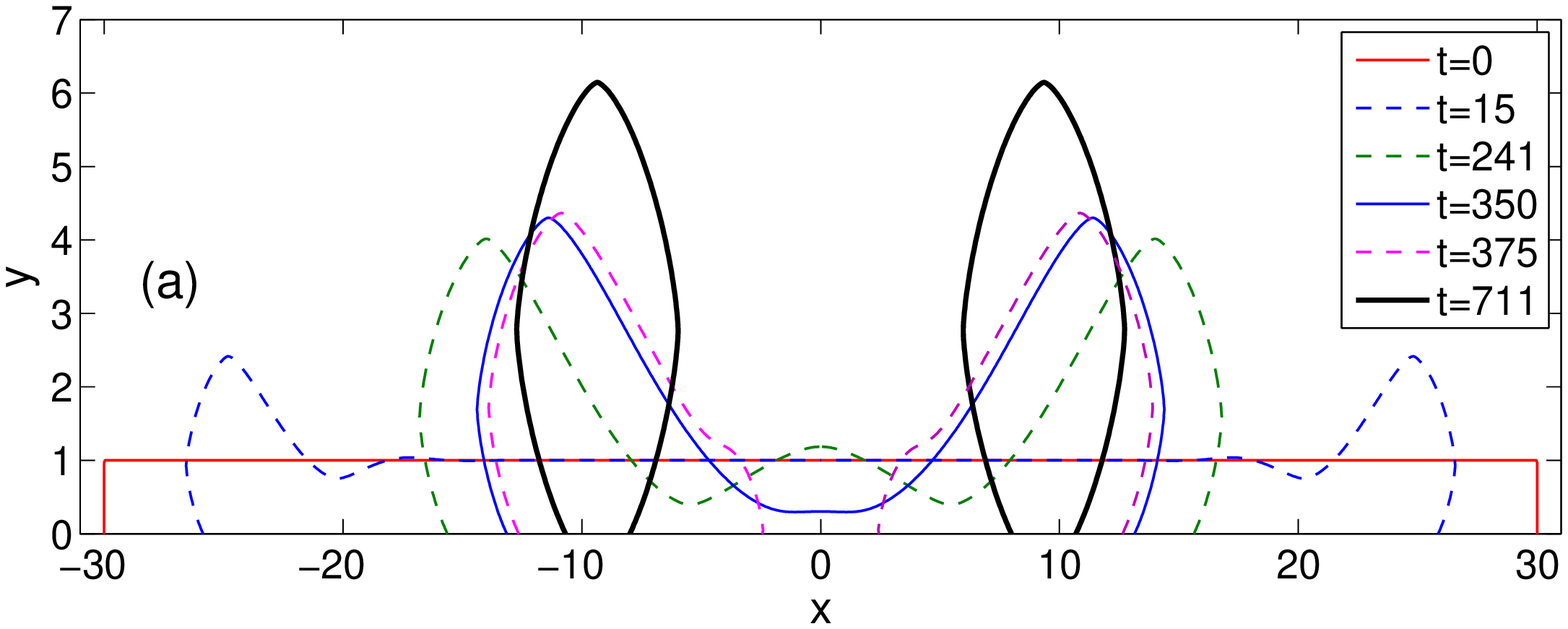}
  \includegraphics[width=0.9\textwidth]{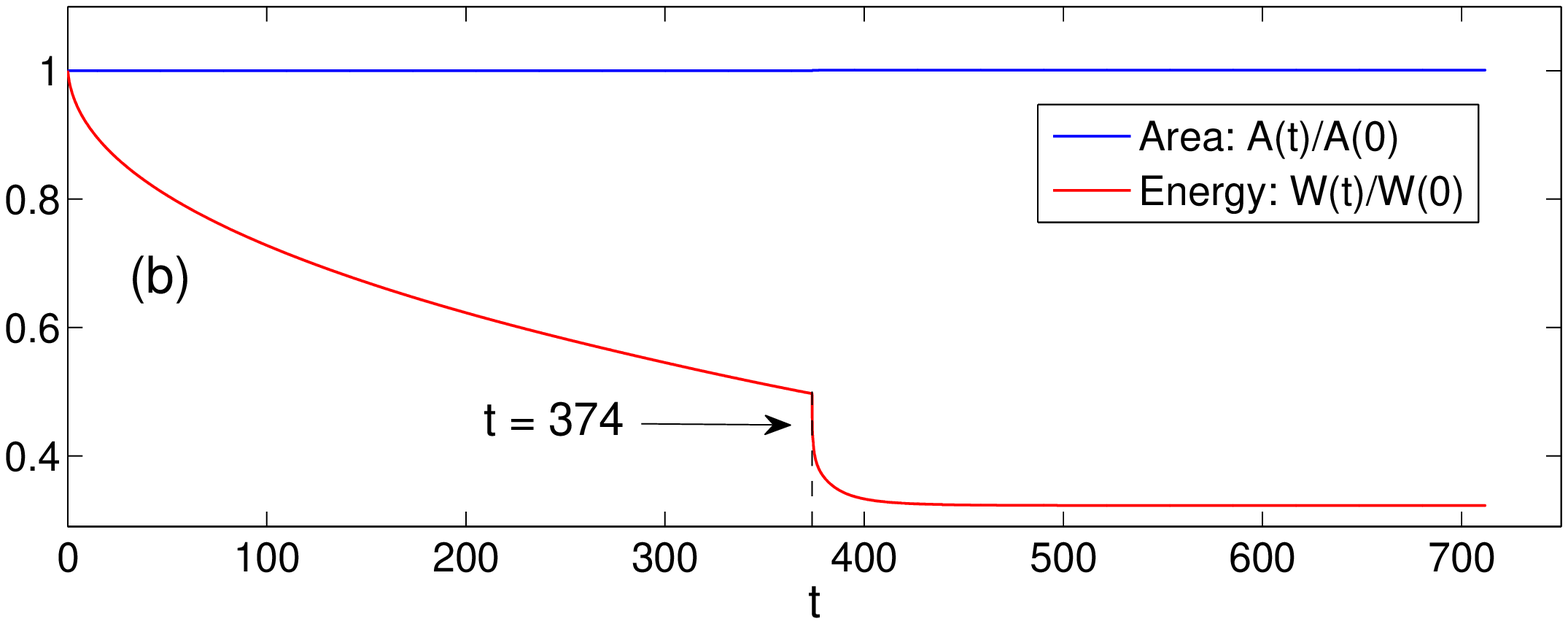}
\caption{(a) The evolution of a  long, thin island (aspect ratio of 60) with weakly anisotropic surface energy ($\beta = 0.06, m = 4, \theta_i = 5\pi/6$). Note  the difference in vertical and horizontal scales. (b) The corresponding temporal evolution of the normalized total free energy and the normalized area (mass). }\label{fig:pinchoff}
\end{figure}

As noted in \cite{Jiang12,Dornel06}, when the aspect ratios of islands are larger than critical values, the islands pinch-off leaving two, three or more islands. Figure~\ref{fig:pinchoff}a shows the temporal evolution of a very large (thin) island (aspect ratio of $60$) with weakly anisotropic surface energy. Figure~\ref{fig:pinchoff}a shows that  surface diffusion kinetics very quickly lead to the formation of ridges at the island edges followed by valleys. As time evolves and the island contact point retracts, these two features become increasing exaggerated, then two valleys merge near the island center. Eventually, the valley at the center of the islands deepens until it touches the substrate, leading to a  pinch-off event  that separates the initial island into a pair of islands. The corresponding evolution of the normalized total free energy and the normalized enclosed area are shown in Fig.~\ref{fig:pinchoff}b.  During the dewetting process, the area (mass) is conserved and the energy decays. The energy undergoes a sharp drop at  $t = 374$ -- the moment when the pinch-off event occurs.

\begin{figure}
  \centering
    \includegraphics[width=0.45\textwidth]{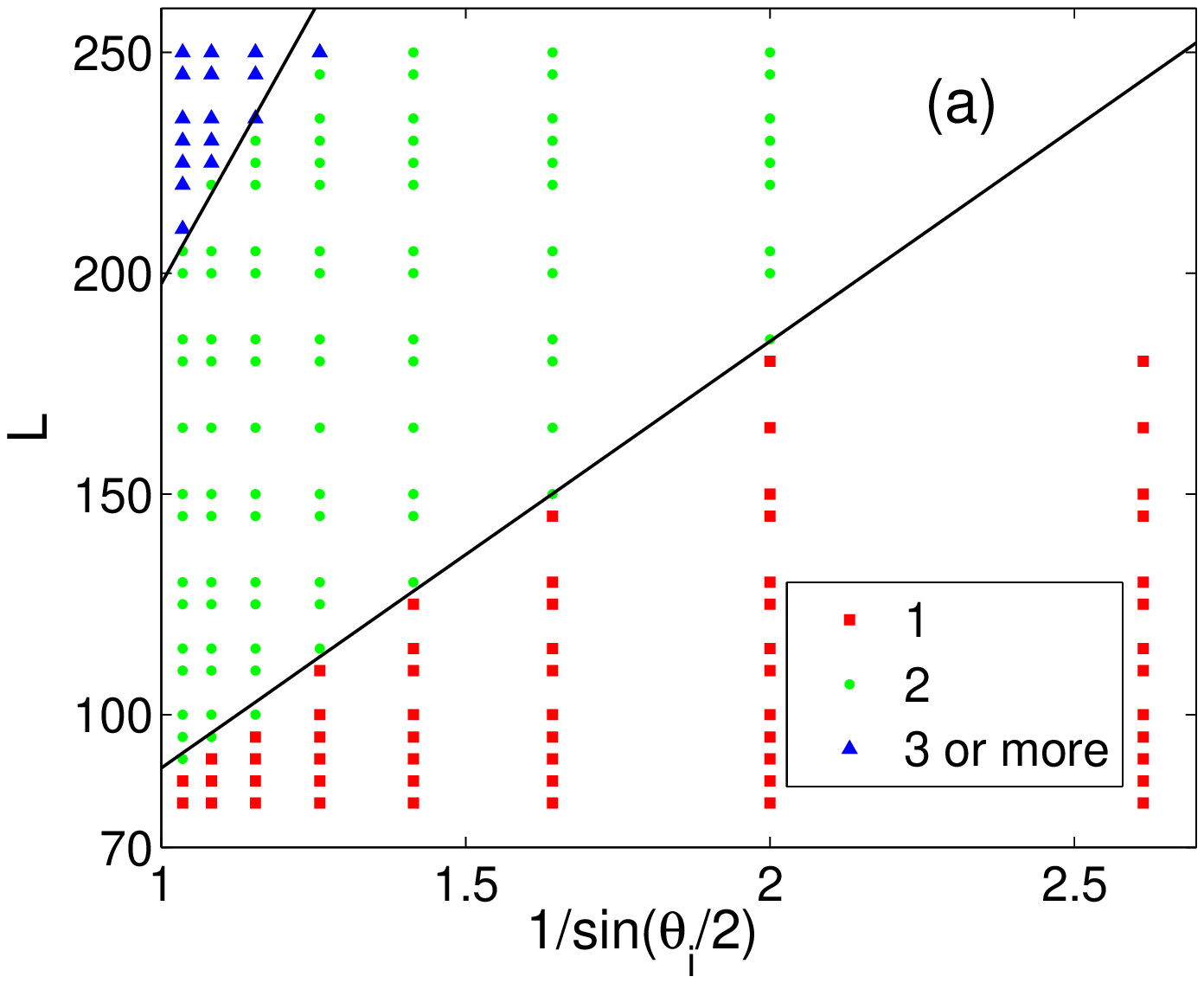}
    \includegraphics[width=0.45\textwidth]{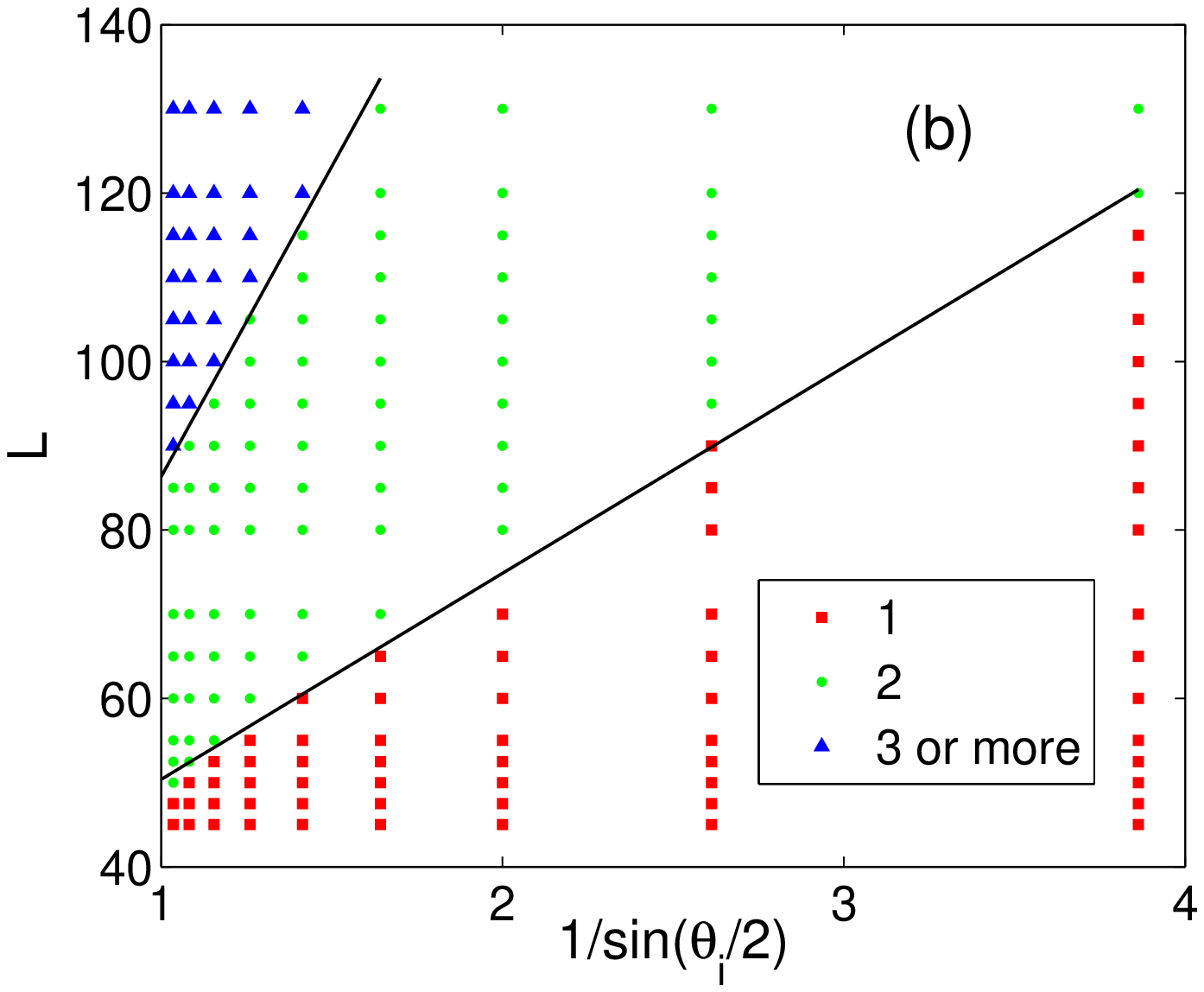}
    \caption{The number of islands formed from the retraction of a high aspect ratio island as a function of initial length $L$ and $\theta_i$ ($h=1$) (a) for the isotropic case and (b) anisotropic case with $\beta = 0.06, m = 4$. In (a), the solid lines are  numerical fits to the results of Dornel, {\it et al.}~\cite{Dornel06}. In (b), the 1-2 islands and 2-3 islands boundaries (solid lines) are  linear fits to the data --- $L = 24.46/\sin(\theta_i/2) + 25.91$ and $L = 73.59/\sin(\theta_i/2) + 12.74$, respectively. }\label{fig:long_agg}
\end{figure}

In addition to the aspect ratio, the parameter $\theta_i$ plays an important role in determining the number of pinch-off events that will occur. We performed a series of numerical simulations for large islands with different aspect ratios and different values of $\theta_i$; the results are shown in Figs.~\ref{fig:long_agg} for both the isotropic case and the weakly anisotropic case and compare these with the results of Dornel, {\it et al.}~\cite{Dornel06}. We observe distinct boundaries between 1, 2 and 3 (or more) islands at late times. For the isotropic case, our results (i.e., the lines that divide between different number of islands - shown in Fig.~\ref{fig:long_agg}a) are consistent with the results of Dornel, {\it et al.}~\cite{Dornel06}. For the anisotropic case, our linear curve fits to identify the 1-2 islands  and 2-3 islands boundaries ($L = 24.46/\sin(\theta_i/2) + 25.91$ and $L = 73.59/\sin(\theta_i/2) + 12.74$, respectively).  Comparing the isotropic and anisotropic results (Figs.~\ref{fig:long_agg}a--b), we see that for the same value of $\theta_i$ an island tends to evolve into a larger number of islands in the anisotropy case than in the isotropic case.
This is in disagreement with the observations of results of Dornel, {\it et al.}~\cite{Dornel06}.

\subsection{Semi-infinite films}

Several earlier studies have shown that a discontinuous film (i.e., a semi-infinite film) retracts such that the distance the contact point moves scales with time according to a power law relation $l \sim t^n$ for sufficiently long time. For the isotropic case, analytical predictions in the small film surface slope limit suggest $n = 1/4$~\cite{Srolovitz86} and $n = 2/5$~\cite{Wong00}. On the other hand, numerical simulations using the sharp interface model~\cite{Wong00} and phase field model~\cite{Jiang12} both suggest that $n\approx0.4$ in the isotropic limit. A study of the anisotropic case~\cite{Kim13,Zucker13} also found $n\approx0.4$ based on numerical simulations with the crystalline model and in experiments on single crystal nickel films on MgO.

\begin{figure}
\centering
\includegraphics[width=0.8\textwidth]{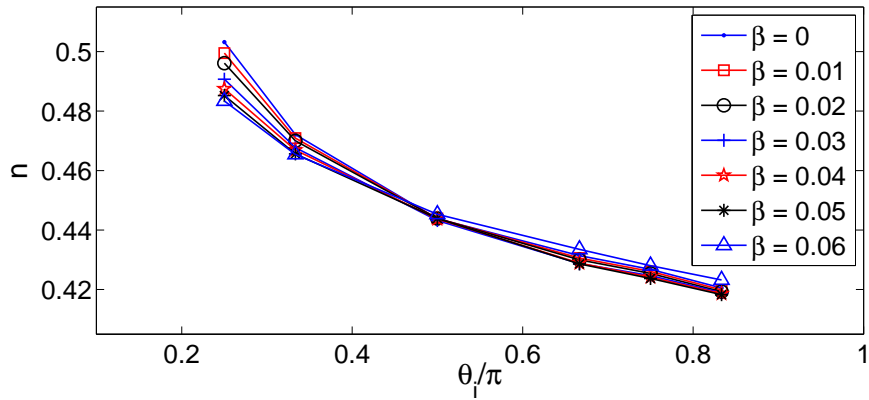}
\caption{The exponent $n$ obtained by fitting the simulation data for the retraction distance of an initially semi-infinite thin film ($l \sim t^n$) versus the corresponding isotropic Young contact angles $\theta_i$ for the case of a weakly anisotropic surface energy with different degrees of anisotropy $\beta$. }\label{fig:pl4}
\end{figure}

We simulated the evolution of a discontinuous film (semi-infinite flat film with a step) with the anisotropic surface energy $\gamma(\theta) = 1+\beta\cos(4\theta)$ and observed a power-law retraction rate. Figure~\ref{fig:pl4} shows the exponent $n$ as a function of $\theta_i$ for different degrees of anisotropy $\beta$.  As shown in Fig.~\ref{fig:pl4}, the power law exponent $n$ are all in the 0.4-0.5 range, depending on $\theta_i$ but nearly independent of the strength of the anisotropy.
The fact that the retraction exponent is sensitive to  $\theta_i$ demonstrates that the retraction exponent is not universal.

\subsection{Infinite films with a  hole}

\begin{figure}
\centering
\includegraphics[width=0.7\textwidth]{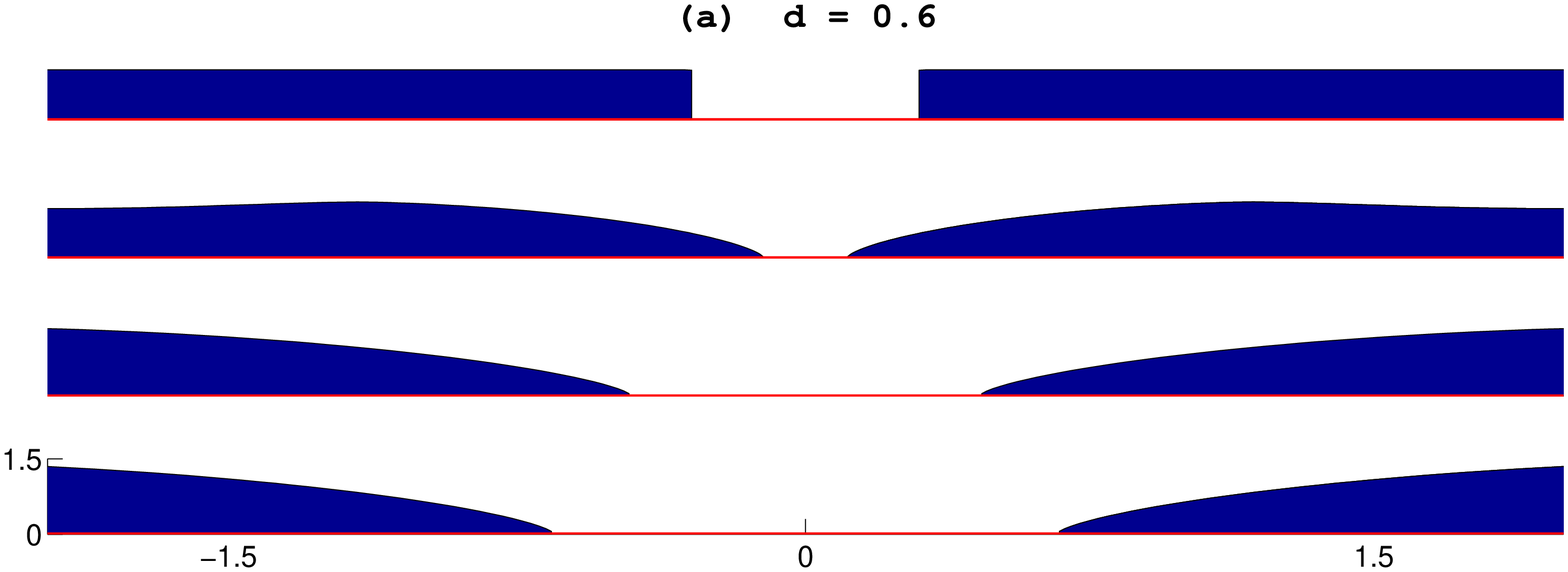}
\includegraphics[width=0.7\textwidth]{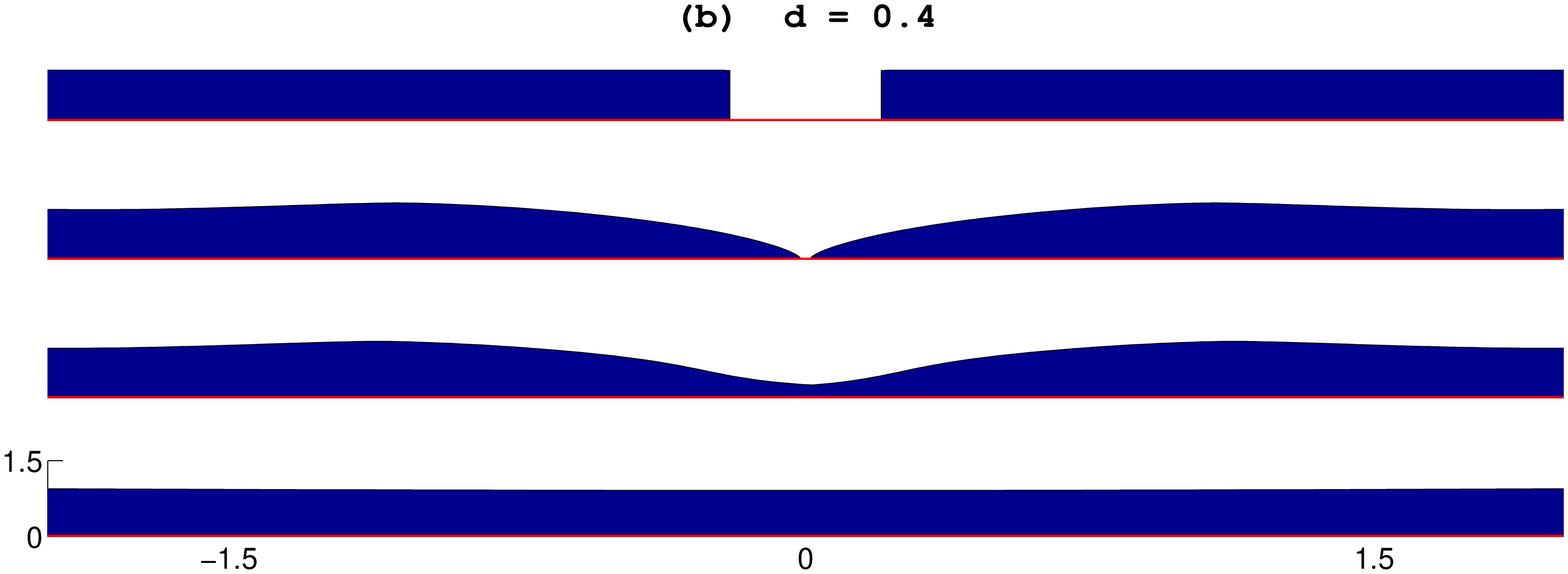}
\includegraphics[width=0.7\textwidth]{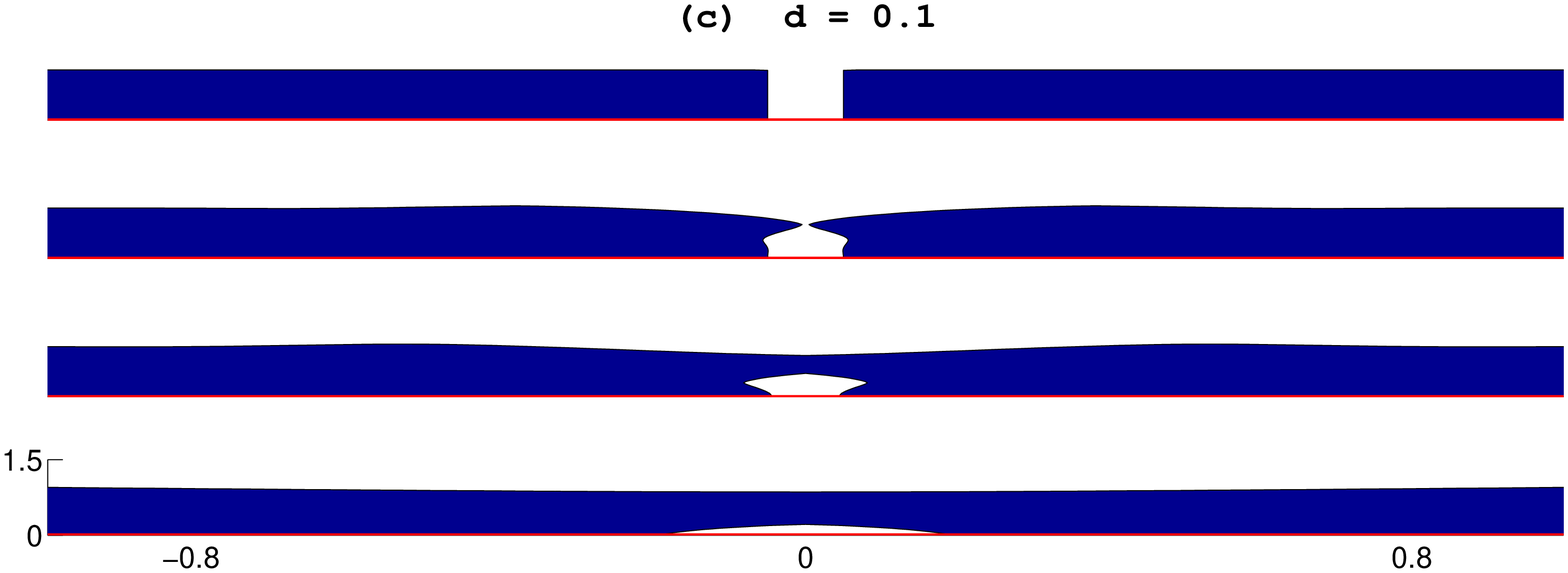}
\caption{Three different types of morphological evolution of an infinite film with a hole of diameter $d$ under anisotropic surface energy conditions, $\gamma(\theta) = 1+0.06\cos(4\theta)$ and $\theta_i = \pi/2$, (a) Case I: dewetting; (b) Case II: wetting; (c) Case III: void. Note that the vertical and horizontal scales are different.}\label{fig:hole_3cases}
\end{figure}

\begin{figure}
\centering
    \includegraphics[width=0.4\textwidth]{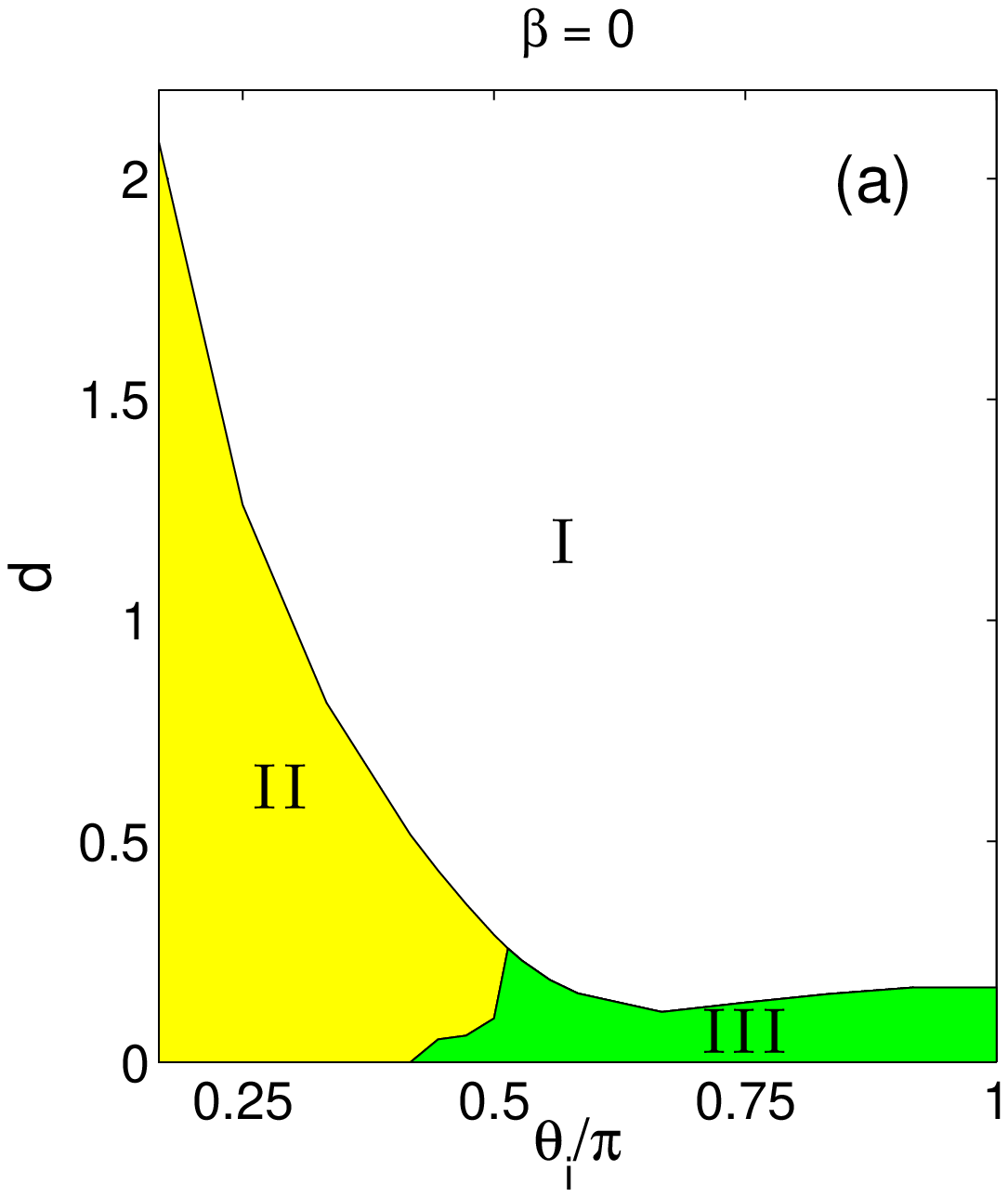}
    \includegraphics[width=0.4\textwidth]{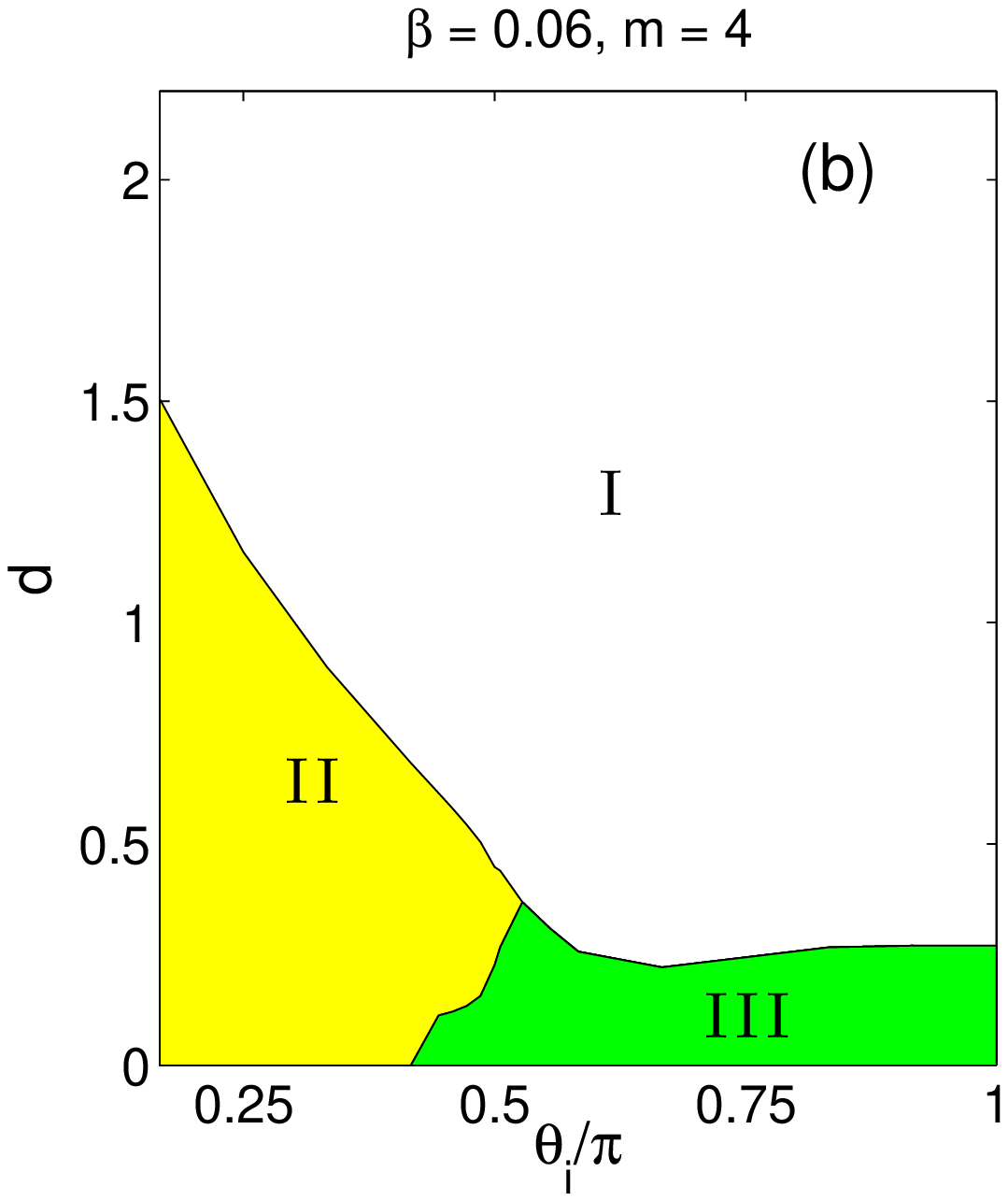}
    \caption{Diagram showing the conditions ($d$ and $\theta_i$) for the occurrence of the three Cases, I-{\sl dewetting}, II-{\sl wetting} and III-{\sl void} for (a) isotropic and (b) anisotropic surface energies (shown by Fig.~\ref{fig:hole_3cases}).}
    \label{fig:holesize_an}
\end{figure}

Finally, we performed numerical simulations for the evolution of an initially continuous thin film containing a single hole from the free surface to the substrate. As reported previously~\cite{Srolovitz86,Jiang12}, there exists a critical hole size above which the the hole gets larger (i.e., Case I - {\sl dewetting}, shown in Fig.~\ref{fig:hole_3cases}a)  or the hole shrinks and closes (Case II - {\sl wetting}, shown in Fig.~\ref{fig:hole_3cases}b).  Interestingly, we find a third case where the two sides of the hole touch and merge, leaving a covered hole/void/bubble at the continuous film-substrate interface   (Case III - {\sl void}, shown in Fig.~\ref{fig:hole_3cases}c). In this case, if $\theta_a<\pi$, the void is stable and of finite extent, but if $\theta_a=\pi$ the void will grow leaving a continuous film disconnected from the substrate.  We note that the case applies for $f(\theta)>0$ for all $\theta$ (see Eq.~(\ref{eqn:rootnumber})).
The occurrence of these three behaviors depends on $\theta_i$ (or $\theta_a$) and the initial size of the hole. Figure~\ref{fig:holesize_an} shows the phase diagram for the relation among the occurrence of the three cases, the parameters $\theta_i$ and the initial hole size $d$ for the isotropic and an anisotropic surface energy cases. As revealed by the figure, it is easier to form a void at the interface (Case III) for thin films with anisotropic surface energy than when the surface energy is isotropic under the same conditions.

\section{Conclusions}

In this paper, we describe a sharp interface model for simulating solid-state dewetting of thin films with weakly anisotropic surface energy. The evolution of the films is governed by  surface diffusion and contact line migration. The derivation of the sharp interface model is based on an energy variational approach. Unlike other sharp interface models, we include a viscosity or finite mobility associated with the moving contact point. This  gives rise to dynamic contact angles that may be different from the equilibrium contact angles from the Young equation. Many observations have been made of dynamic triple junction angles in grain boundary migration and contact line angles in liquid wetting of substrates that differ markedly from static equilibrium contact angles.  When the surface energy is anisotropic, the Young equation has multiple roots.  With the finite contact line mobility, the system naturally finds the correct roots at all times during the morphology evolution.
We proposed a numerical approach, based upon an explicit finite difference scheme combined with the cubic spline interpolation for evolving marker points, for solving the sharp interface model. Numerical results for solid-state dewetting in two dimensions demonstrate the excellent performance of the method, including stability, convergence and numerical efficiency.

With the validated mathematical and numerical approaches, we simulated the evolution of thin film islands, semi-infinite films, and films with  holes as a function of film dimensions, Young's angle $\theta_i$, anisotropy strength and symmetry, and film crystal orientation relative to the substrate normal. Like others, we found that contact point retraction rate can be well described by a power-law,  $l \sim t^n$.  Our results demonstrate that the exponent $n$ is not universal; it is sensitive to the Young's angle $\theta_i$ (and insensitive to anisotropy).  We have also observed that in addition to classical wetting (where holes in a film heal) and dewetting (where holes in a film grow), another possibility is where the holes heal leaving a continuous film but with a void at the film/substrate interface which can of finite or infinite extent. Surface energy anisotropy was also shown to (i) increase the instability that leads to island break-up into multiple islands, (ii) enhance hole healing, and (iii) lead to finite island size even under some conditions where the Young's angle $\theta_i$ suggests that the film wets the substrate. The numerical results presented in the paper capture many of the complexities associated with solid-state dewetting experiments~\cite{Thompson12,Ye10a,Ye10b,Ye11a,Ye11b,Kim13}.

While, in the present model, we focused on the weakly anisotropic case, the model is readily generalizable to the strongly anisotropic case. In the strongly anisotropic case, the anisotropic evolution equation \eqref{eqn:variation4}  becomes ill-posed; this can be simply overcome by adopting a regularization  approach~\cite{Carlo92,Eggleston01,Torabi10} to make it  well-posed and/or incorporating a more  advanced numerical technique \cite{Hou94,Barrett11,Barrett12} that leads to efficient implementations. The relaxation/finite contact point mobility approach, developed above, and be used to replace the contact line boundary  conditions in the strongly anisotropic case.

\clearpage

\appendix{
\section{Derivation of the first variation of the free energy}
Denoting $\delta$ as the variation of a functional, applying it to variations of the free energy, $W$, in Eq.~\eqref{eqn:2Denergy}, and using the integration by parts, we get
\begin{eqnarray}
\delta W &=& \int_{x_c^l}^{x_c^r}\left[(1+h_x^2)^{1/2}\frac{d\widetilde{\gamma}}{dh_x}+\frac{\widetilde{\gamma} h_x}{(1+h_x^2)^{1/2}}\right]
(\delta h_x)~dx \nonumber \\
&& + \left[\widetilde{\gamma}(1+h_x^2)^{1/2}+\gamma_{\scriptscriptstyle {FS}}-\gamma_{\scriptscriptstyle {VS}}\right]_{x=x_c^r}
\delta x_c^r
- \left[\widetilde{\gamma}(1+h_x^2)^{1/2}+\gamma_{\scriptscriptstyle {FS}}-\gamma_{\scriptscriptstyle {VS}}\right]_{x=x_c^l}
\delta x_c^l, \nonumber\\
&=&-\int_{x_c^l}^{x_c^r}\left[(1+h_x^2)^{1/2}h_{xx}\frac{d^2\widetilde{\gamma}}{dh_x^2}+
\frac{2h_xh_{xx}}{(1+h_x^2)^{1/2}}\frac{d\widetilde{\gamma}}{dh_x}
+\frac{\widetilde{\gamma} h_{xx}}{(1+h_x^2)^{3/2}}\right](\delta h)~dx   \nonumber \\
&&+\left[(1+h_x^2)^{1/2}\frac{d\widetilde{\gamma}}{dh_x}+\frac{\widetilde{\gamma} h_x}{(1+h_x^2)^{1/2}}\right]_{x=x_c^r}(\delta h)|_{x=x_c^r}
  \nonumber  \\
&&-\left[(1+h_x^2)^{1/2}\frac{d\widetilde{\gamma}}{dh_x}+\frac{\widetilde{\gamma} h_x}{(1+h_x^2)^{1/2}}\right]_{x=x_c^l}(\delta h)|_{x=x_c^l}
  \nonumber\\
&&+\left[\widetilde{\gamma}(1+h_x^2)^{1/2}+\gamma_{\scriptscriptstyle {FS}}-\gamma_{\scriptscriptstyle {VS}}\right]_{x=x_c^r}
\delta x_c^r -\left[\widetilde{\gamma}(1+h_x^2)^{1/2}+\gamma_{\scriptscriptstyle {FS}}-\gamma_{\scriptscriptstyle {VS}}\right]_{x=x_c^l}
\delta x_c^l \nonumber\\
&=&-\int_{x_c^l}^{x_c^r}\left[(1+h_x^2)^{1/2}h_{xx}\frac{d^2\widetilde{\gamma}}{dh_x^2}
+\frac{2h_xh_{xx}}{(1+h_x^2)^{1/2}}\frac{d\widetilde{\gamma}}{dh_x}
+\frac{\widetilde{\gamma} h_{xx}}{(1+h_x^2)^{3/2}}\right](\delta h)~dx   \nonumber \\
&&+\left[\frac{\widetilde{\gamma}}{(1+h_x^2)^{1/2}}-h_x(1+h_x^2)^{1/2}\frac{d\widetilde{\gamma}}{d h_x}+\gamma_{\scriptscriptstyle {FS}}-\gamma_{\scriptscriptstyle {VS}}\right]_{x=x_c^r}\delta x_c^r   \nonumber \\
&&-\left[\frac{\widetilde{\gamma}}{(1+h_x^2)^{1/2}}-h_x(1+h_x^2)^{1/2}\frac{d\widetilde{\gamma}}{d h_x}+\gamma_{\scriptscriptstyle {FS}}-\gamma_{\scriptscriptstyle {VS}}\right]_{x=x_c^l}\delta x_c^l,
\label{eqn:step3}
\end{eqnarray}
where $\delta x_c^r$ and $\delta x_c^l$ denote the variations of the free energy with respect to the positions of the right and left contact points and we have used
$(\delta h)|_{x=x_c^r}=-h_x|_{x=x_c^r}\delta x_c^r$ and $(\delta h)|_{x=x_c^l}=-h_x|_{x=x_c^l}\delta x_c^l$.
Making use of these results, we find the first variation of the total energy functional with respect to $h(x), x_c^r, x_c^l$:
\begin{equation}
\frac{\delta W}{\delta h}=-\left[(1+h_x^2)^{1/2}h_{xx}\frac{d^2\widetilde{\gamma}}{dh_x^2}+\frac{2h_xh_{xx}}
{(1+h_x^2)^{1/2}}\frac{d\widetilde{\gamma}}{dh_x}
+\frac{\widetilde{\gamma} h_{xx}}{(1+h_x^2)^{3/2}}\right], \qquad x \in (x_c^l, x_c^r),
\label{eqn:variation1}
\end{equation}

\begin{equation}
\frac{\delta W}{\delta x_c^r}=\left[\frac{\widetilde{\gamma}}{(1+h_x^2)^{1/2}}-h_x(1+h_x^2)^{1/2}\frac{d\widetilde{\gamma}}{d h_x}+\gamma_{\scriptscriptstyle {FS}}-
\gamma_{\scriptscriptstyle {VS}}\right]_{x=x_c^r},
\label{eqn:variation2}
\end{equation}

\begin{equation}
\frac{\delta W}{\delta x_c^l}=-\left[\frac{\widetilde{\gamma}}{(1+h_x^2)^{1/2}}-h_x(1+h_x^2)^{1/2}\frac{d\widetilde{\gamma}}{d h_x}+\gamma_{\scriptscriptstyle {FS}}-
\gamma_{\scriptscriptstyle {VS}}\right]_{x=x_c^l}.
\label{eqn:variation3}
\end{equation}
Introducing the surface normal angle $\theta$, then we immediately have the following relations:
\begin{equation}
\cos \theta = \frac{1}{(1+h_x^2)^{1/2}},
\label{eqn:theta1}
\end{equation}
\begin{equation}
\widetilde{\gamma}\,'(\theta)=(1+h_x^2)\frac{d \widetilde{\gamma}}{d h_x},
\label{eqn:theta2}
\end{equation}
\begin{equation}
\widetilde{\gamma}\,''(\theta)=(1+h_x^2)^2\frac{d^2 \widetilde{\gamma}}{d h_x^2}+2h_x(1+h_x^2)\frac{d \widetilde{\gamma}}{d h_x},
\label{eqn:theta3}
\end{equation}
\begin{equation}
\kappa = -\frac{h_{xx}}{(1+h_x^2)^{3/2}}.
\label{eqn:kappa}
\end{equation}
Inserting \eqref{eqn:theta1}-\eqref{eqn:kappa} into Eqs.~\eqref{eqn:variation1}-\eqref{eqn:variation3},
we obtain the variations used in Eqs.~\eqref{eqn:variation4}-\eqref{eqn:variation6}.

\section{Winterbottom construction}

In this paper, we use the following form of the chemical potential for the weakly anisotropic surface energy:
\[
  \mu = \big[\gamma(\theta) + \gamma\,''(\theta)\big]\kappa, \qquad \gamma(\theta) = 1 + \beta \cos[m(\theta+\phi)].
\]
For simplicity, we assume here that $\phi = 0$. The corresponding Wulff shape
(the equilibrium crystal shape without a  substrate) for this surface energy can be explicitly written as~\cite{Eggleston01}:
\[
  \begin{cases}
    x(\theta) = -\gamma(\theta)\sin\theta - \gamma\,'(\theta)\cos\theta,\\
    y(\theta) = \gamma(\theta)\cos\theta - \gamma\,'(\theta)\sin\theta,
  \end{cases}  \qquad  \theta\in [-\pi, \pi].
\]
We note a slight difference in notation as compared to that in~\cite{Eggleston01}; in \cite{Eggleston01}, $\theta$ is the angle between the surface outer normal and the $x$-axis while here $\theta$ is the angle between the surface outer normal and the $y$-axis.

The equilibrium shape for an island  on a flat, rigid substrate can be constructed, using the Winterbottom approach~\cite{winterbottom67}, by adding a substrate parallel to the $x$-axis to the  Wulff shape (above). The distance from the substrate to the Wulff point (center of the Wulff shape) is $\abs{\cos\theta_i}$, i.e. $\cos\theta_i=(\gamma_{\scriptscriptstyle {VS}}-\gamma_{\scriptscriptstyle {FS}})/\gamma_0 \in [-1,1]$. More precisely, if $\theta_i \in (\pi/2, \pi]$, the Wulff point lies within the Wulff shape and the substrate is at $y=\cos\theta_i$ and parallel to the $x$-axis; if $\theta_i \in [0,\pi/2)$, the Wulff point lies outside of the equilibrium shape and the substrate is at  $y=\cos\theta_i$ and parallel to the $x$-axis. Therefore, the equilibrium shape of a crystal on a flat substrate which is coincident with the $x$-axis (i.e., at $y=0$) can be rewritten as:
\[
  \begin{cases}
    x(\theta) = -\gamma(\theta)\sin\theta - \gamma\,'(\theta)\cos\theta,\\
    y(\theta) = \gamma(\theta)\cos\theta - \gamma\,'(\theta)\sin\theta - \cos\theta_i,
  \end{cases}  \qquad y\ge 0 \quad (\mbox{or}~\theta \in [-\theta_a, \theta_a]),
\]
where the equilibrium contact angle $\theta_a \in [0,\pi]$ is determined from
\[
  \cos(\theta_a) = \frac{(-y\,'(\theta),x\,'(\theta))\cdot(0,1)}
  {\sqrt{(x\,'(\theta))^2 + (y\,'(\theta))^2}}\bigg|_{y = 0} = \cos(\theta)\bigg|_{y = 0}.
\]
From this expression, we see that the equilibrium contact angle $\theta_a$ can be any root of the equation:
\[
  \gamma(\theta)\cos\theta - \gamma\,'(\theta) \sin\theta - \cos\theta_i = 0,
\]
which satisfies the contact angle equation~(\ref{eqn:forcebalance}).

Finally, we ensure  conservation of the area (mass) of the crystal by a normalization procedure. Assume that the enclosed area of the equilibrium shape given by the  Winterbottom construction, above, is $A_w$, i.e. $A_w = -\int_{-\theta_a}^{\theta_a}
y(\theta)\, x\,'(\theta)\, \mathrm{d}\theta$, then the normalized equilibrium shape is explicitly given by:
\[
  \begin{cases}
    x_e(\theta) = -\sqrt{\frac{A_0}{A_w}}\big(\big[1+\beta\cos(m\theta)\big]\sin\theta - m\beta\sin(m\theta) \cos\theta\big),\\
    y_e(\theta) =\sqrt{\frac{A_0}{A_w}}\big(\big[1+\beta\cos(m\theta)\big]\cos\theta + m\beta\sin(m\theta) \sin\theta - \cos\theta_i\big),
  \end{cases} \quad \theta\in [-\theta_a, \theta_a],
\]
where $A_0$ is the area (or mass) enclosed by the film and substrate.
}

\begin{acknowledgments}
This work was supported by the National Science Foundation of China (W.J.) and
the Singapore A*STAR SERC PSF-grant 1321202067 (Y.W. and W.B.).
Part of the work was done when the authors
were visiting Beijing Computational Science Research Center in 2014.
\end{acknowledgments}



\begin{thebibliography}{1}%
\makeatletter
\providecommand \@ifxundefined [1]{%
 \@ifx{#1\undefined}
}%
\providecommand \@ifnum [1]{%
 \ifnum #1\expandafter \@firstoftwo
 \else \expandafter \@secondoftwo
 \fi
}%
\providecommand \@ifx [1]{%
 \ifx #1\expandafter \@firstoftwo
 \else \expandafter \@secondoftwo
 \fi
}%
\providecommand \natexlab [1]{#1}%
\providecommand \enquote  [1]{``#1''}%
\providecommand \bibnamefont  [1]{#1}%
\providecommand \bibfnamefont [1]{#1}%
\providecommand \citenamefont [1]{#1}%
\providecommand \href@noop [0]{\@secondoftwo}%
\providecommand \href [0]{\begingroup \@sanitize@url \@href}%
\providecommand \@href[1]{\@@startlink{#1}\@@href}%
\providecommand \@@href[1]{\endgroup#1\@@endlink}%
\providecommand \@sanitize@url [0]{\catcode `\\12\catcode `\$12\catcode
  `\&12\catcode `\#12\catcode `\^12\catcode `\_12\catcode `\%12\relax}%
\providecommand \@@startlink[1]{}%
\providecommand \@@endlink[0]{}%
\providecommand \url  [0]{\begingroup\@sanitize@url \@url }%
\providecommand \@url [1]{\endgroup\@href {#1}{\urlprefix }}%
\providecommand \urlprefix  [0]{URL }%
\providecommand \Eprint [0]{\href }%
\providecommand \doibase [0]{http://dx.doi.org/}%
\providecommand \selectlanguage [0]{\@gobble}%
\providecommand \bibinfo  [0]{\@secondoftwo}%
\providecommand \bibfield  [0]{\@secondoftwo}%
\providecommand \translation [1]{[#1]}%
\providecommand \BibitemOpen [0]{}%
\providecommand \bibitemStop [0]{}%
\providecommand \bibitemNoStop [0]{.\EOS\space}%
\providecommand \EOS [0]{\spacefactor3000\relax}%
\providecommand \BibitemShut  [1]{\csname bibitem#1\endcsname}%
\let\auto@bib@innerbib\@empty
\bibitem [{Note1()}]{Note1}%
  \BibitemOpen
  \bibinfo {note} {For simplicity of presentation, we assume $h(x)$ is a
  single-valued function of $x$. When $h(x)$ is a multi-valued function, this
  procedure is applied using an arc length parameterization.}\BibitemShut
  {Stop}%
\end{thebibliography}%


\begin{thebibliography}{99}
{
  \bibitem{Thompson12}
  C.V.~Thompson, Annu. Rev. Mater. Res. 42, 399 (2012).

  \bibitem{Jiran90}
  E.~Jiran, C.V.~Thompson, J. Electron Mater. 19, 1153 (1990).

  \bibitem{Jiran92}
  E.~Jiran, C.V.~Thompson, Thin Solid Films 208, 23 (1992).

  \bibitem{Ye10a}
  J.~Ye, C.V.~Thompson, Appl. Phys. Lett. 97, 071904 (2010).

  \bibitem{Ye10b}
  J.~Ye, C.V.~Thompson, Phys. Rev. B 82, 193408 (2010).

  \bibitem{Ye11a}
  J.~Ye, C.V.~Thompson, Acta Mater. 59, 582 (2011).

  \bibitem{Ye11b}
  J.~Ye, C.V.~Thompson, Adv. Mater. 23, 1567 (2011).

  \bibitem{Jiang12}
  W.~Jiang, W.~Bao, C.V.~Thompson, D.J.~Srolovitz, Acta Mater. 60, 5578 (2012).

  \bibitem{Kim13}
  G.H.~Kim, R.V.~Zucker, W.C.~Carter, C.V.~Thompson, J. Appl. Phys. 113, 043512 (2013).

  \bibitem{Zucker13}
  R.V.~Zucker, J.H.~Kim, W.C.~Carter, C.V.~Thompson, Comptes Rendus Physique 14, 564 (2013).

  \bibitem{Rabkin11}
  L.~Klinger, D.~Amram, E.~Rabkin, Scripta Mater. 64, 962 (2011).

  \bibitem{Rabkin14a}
  A.~Kosinova, L.~Klinger, O.~Kovalenko, E.~Rabkin, Scripta Mater. 82, 33 (2014).

  \bibitem{Rabkin14b}
  E.~Rabkin, D.~Amram, E.~Alster, Acta Mater. 74, 30 (2014).

  \bibitem{Pierre11}
  M.~Dufay, O.~Pierre-Louis, Phys. Rev. Lett. 106, 105506 (2011).

  \bibitem{Qian06}
  T.Z.~Qian, X.P.~Wang, P.~Sheng, J. Fluid Mech. 564, 333 (2006).

  \bibitem{Ren07}
  W.~Ren, W.~E, Phys. Fluid 19, 022101 (2007).

  \bibitem{Ren10}
  W.~Ren, D.~Hu, W.~E, Phys. Fluid 22, 102103 (2010).

  \bibitem{Ren11}
  W.~Ren, W.~E, Phys. Fluid 23, 072103 (2011).

  \bibitem{Srolovitz86}
  D.J.~Srolovitz, S.A.~Safran, J. Appl. Phys. 60, 255 (1986).

  \bibitem{Wong00}
  H.~Wong, P.W.~Voorhees, M.J.~Miksis, S.H.~Davis, Acta Mater. 48, 1719 (2000).

  \bibitem{Du10}
  P.~Du, M.~Khenner, H.~Wang, J. Comput. Phys. 229, 813 (2010).

  \bibitem{Dornel06}
  E.~Dornel, J.C.~Barbe, F.~Crecy, G.~Lacolle, J.~Eymery, Phys. Rev. B, 73, 115427 (2006).

  \bibitem{Carter95}
  W.C.~Carter, A.R.~Roosen, J.W. Cahn, J.E.~Taylor, Acta Metall. Mater. 43, 4309 (1995).

  \bibitem{Eggleston01}
  J.J.~Eggleston, G.B.~McFadden, P.W.~Voorhees, Physica D 150, 91 (2001).

  \bibitem{Min06}
  D.H.~Min, H.~Wong, J. Appl. Phys. 100, 053523 (2006).

  \bibitem{Czubayko98}
  U.~Czubayko, V.G.~Sursaeva, G.~Gottstein, L.S.~Shvindlerman, Acta Mater. 46, 5863 (1998).

  \bibitem{Upmanyu02}
  M.~Upmanyu, D.J.~Srolovitz, L.S.~Shvindlerman, G.~Gottstein, Acta Mater. 50, 1405 (2002).

  \bibitem{deGennes85}
  P.~de Gennes, Rev. Mod. Phys. 57, 827 (1985).

  \bibitem{Ralston08}
  J.~Ralston, M.~Popescu, R.~Seder, Annu. Rev. Mater. Res. 38, 23 (2008).

  \bibitem{winterbottom67}
  W.L.~Winterbottom, Acta Matall. 15, 303 (1967).

 \bibitem{Carlo92}
 A.~Carlo, M.~Gurtin, P.~Podio-Guidugli, SIAM J. Appl. Math. 52, 1111 (1992).

 \bibitem{Torabi10}
 S.~Torabi, J.S.~Lowengrub, A.~Voigt, S.~Wise, Proc. R. Soc. A 465, 1337 (2010).

 \bibitem{Hou94}
 T.Y.~Hou, J.S.~Lowengrub, M.J.~Shelley, J. Comput. Phys. 114, 312 (1994).

  \bibitem{Barrett11}
  J.W.~Barrett, H.~Garcke, R.~Numberg, Numer. Methods Partial. Differ. Equ. 27, 1 (2011).

  \bibitem{Barrett12}
  J.W.~Barrett, H.~Garcke, R.~Numberg, Numer. Math. 120, 489 (2012).

}
\end{thebibliography}
\end{document}